\documentclass[iop]{emulateapj}
\usepackage{apjfonts}
\newcommand{\be}{\begin{equation}}
\newcommand{\ee}{\end{equation}}
\newcommand{\Msun}{M_{\odot}}
\def\kms{\ {\rm km\, s}^{-1}}
\def\teff{T_{\rm eff}}

\def\mlk{\ M/L_K}
\def\mli{\ M/L_I}

\shortauthors{CONROY \& VAN DOKKUM}
\shorttitle{The Stellar IMF in Early-Type Galaxies. II. Results}

\begin{document}

\title{The Stellar Initial Mass Function in Early-Type Galaxies From
  Absorption Line Spectroscopy. II. Results}

\author{Charlie Conroy\altaffilmark{1,2}
  \& Pieter G. van Dokkum \altaffilmark{3}}

\altaffiltext{1}{Harvard-Smithsonian Center for Astrophysics,
  Cambridge, MA, USA} 
\altaffiltext{2}{Department of Astronomy \& Astrophysics, University
  of California, Santa Cruz, CA, USA}
\altaffiltext{3}{Department of Astrophysical
  Sciences, Yale University, New Haven, CT, USA}

\slugcomment{Accepted for publication in ApJ}

\begin{abstract}

  The spectral absorption lines in early-type galaxies contain a
  wealth of information regarding the detailed abundance pattern, star
  formation history, and stellar initial mass function (IMF) of the
  underlying stellar population.  Using our new population synthesis
  model that accounts for the effect of variable abundance ratios of
  11 elements, we analyze very high quality absorption line spectra of
  38 early-type galaxies and the nuclear bulge of M31.  These data
  extend to $1\mu m$ and they therefore include the IMF-sensitive
  spectral features NaI, CaII, and FeH at $0.82\mu m$, $0.86\mu m$ and
  $0.99\mu m$, respectively.  The models fit the data well, with
  typical rms residuals $\lesssim1\%$.  Strong constraints on the IMF
  and therefore the stellar mass-to-light ratio, $(M/L)_{\rm stars}$,
  are derived for individual galaxies.  We find that the IMF becomes
  increasingly bottom-heavy with increasing velocity dispersion and
  [Mg/Fe].  At the lowest dispersions and [Mg/Fe] values the derived
  IMF is consistent with the Milky Way IMF, while at the highest
  dispersions and [Mg/Fe] values the derived IMF contains more
  low-mass stars (is more bottom-heavy) than even a Salpeter IMF.  Our
  best-fit $(M/L)_{\rm stars}$ values do not exceed dynamically-based
  $M/L$ values.  We also apply our models to stacked spectra of four
  metal-rich globular clusters in M31 and find an $(M/L)_{\rm stars}$
  that implies {\it fewer} low-mass stars than a Milky Way
  \citep{Kroupa01} IMF, again agreeing with dynamical constraints.  We
  discuss other possible explanations for the observed trends and
  conclude that variation in the IMF is the simplest and most
  plausible.

\end{abstract}



\section{Introduction}
\label{s:intro}

The stellar IMF plays a central role in many areas of astrophysics.
It sets the overall stellar mass scale of galaxies, determines the
amount of energetic feedback following an episode of star formation
via the ratio of high-to-low mass stars, governs the nucleosynthetic
history of galaxies, and, more fundamentally, provides insight into
the physics of star formation.

Despite the central importance of the IMF, a comprehensive physical
theory for its origin and variation with environment does not yet
exist.  It has been argued that the characteristic mass scale and
shape of the IMF is set by the Jeans mass \citep[e.g.,][]{Larson98,
  Larson05}, by feedback from protostars \citep{Silk95, Adams96,
  Krumholz11}, or by the distribution of densities in a supersonically
turbulent interstellar medium \citep{Padoan97, Padoan02, Hennebelle08,
  Hopkins12a, Hopkins12b}.  

Despite the variety of ideas, every theory predicts at least some
variation in the IMF with physical properties.  Remarkably,
observations in our Galaxy of star forming regions, open and globular
clusters, and field stars have found little variation in the IMF
\citep[e.g.,][]{Scalo86, Kroupa01, Bastian10, Kroupa12}.  The
observations are however complicated by a myriad of selection effects,
biases, and correction factors.  Observed luminosity functions must be
corrected for binarity, stellar evolution, and, for globular clusters,
dynamical evolution that preferentially ejects low-mass stars from the
clusters.  Moreover, direct constraints on the low-mass
($M\lesssim0.5\Msun$) IMF are limited to relatively mundane
environments. Nonetheless, unambiguous evidence for IMF variation from
direct star counts does not currently exist.

In nearby and distant galaxies, direct estimates of the IMF to
$\approx0.1\Msun$ from star counts is currently impossible, and,
except perhaps for a handful of Local Group galaxies, will remain
impossible for the indefinite future.  Less direct probes of the IMF
are therefore required.

\begin{deluxetable*}{lccl}
\tablecaption{Model Parameters}
\tablehead{ \colhead{Parameter} &\colhead{Prior} & 
\colhead{Units} &\colhead{Notes} }
\startdata
$v_z$      & (-1,000,10,000) & $\kms$ & recession velocity \\
$\sigma$   & (20,400)        & $\kms$ & velocity dispersion\\
$[$Fe/H$]$ & (-0.4,0.4)      &        & Iron abundance \\
$[$O,Ne,S/Fe$]$ & (-0.4,0.6) &        & Oxygen, Neon, Sulfur abundance \\
$[$C/Fe$]$ & (-0.4,0.4)      &        & Carbon abundance \\
$[$N/Fe$]$ & (-0.4,0.8)      &        & Nitrogen abundance \\
$[$Na/Fe$]$ & (-0.4,1.3)     &        & Sodium abundance \\
$[$Mg/Fe$]$ & (-0.4,0.6)     &        & Magnesium abundance \\
$[$Si/Fe$]$ & (-0.4,0.4)     &        & Silicon abundance \\
$[$Ca/Fe$]$ & (-0.4,0.4)     &        & Calcium abundance \\
$[$Ti/Fe$]$ & (-0.4,0.4)     &        & Titanium abundance \\
$[$Cr/Fe$]$ & (-0.4,0.4)     &        & Chromium abundance \\
$[$Mn/Fe$]$ & (-0.4,0.4)     &        & Manganese abundance \\
age         & (4,15.0)       & Gyr    & Age of bulk population\\
log$(f_y)$ & (-5.0,-0.3)     & & fraction of young (3 Gyr) stars \\
$\alpha_1$ & (0.0,3.5) & & IMF slope over $0.1\Msun<M<0.5\Msun$ \\
$\alpha_2$ & (0.0,3.5) & & IMF slope over $0.5\Msun<M<1.0\Msun$ \\
$\alpha_3$ & 2.3 & & IMF slope over $1.0\Msun<M<100\Msun$ \\
$\Delta(\teff)$ & (-50,50)  & $K$  & temperature offset applied to all stars \\
log(M7III) & (-5.0,-0.3)  &    & fraction of additional M7III light \\
log$(f_{\rm hot})$ & (-5.0,-0.3)&  & fraction of additional hot stars \\
$T_{\rm hot}$ & (1,3)     & $10^4\,K$ & Temperature of additional hot stars
\enddata
\vspace{0.1cm} 
\tablecomments{The prior is flat within the range defined in the table
  and cuts off sharply outside of the prior range.  The parameter
  $\alpha_3$ is fixed to the value 2.3.}
\label{t:params}
\end{deluxetable*}

It was recognized in the 1960s that low-mass stars, though
individually faint, can be detected in the integrated light of
early-type galaxies \citep{Spinrad62}.  Early attempts to exploit this
fact to measure the IMF in early-type galaxies lead to a variety of
conflicting claims \citep{Spinrad71, Cohen78, Faber80, Hardy88,
  Couture93, Carter86, Delisle92, Couture93}, owing largely to the
inadequate quality of the data and models.  Thanks largely to the
increased quality of the stellar interior and atmospheric models and
data in the intervening years, this technique is now capable of
providing powerful constraints on the IMF \citep[e.g.,][]{Cenarro03}.

In recent work we built a new population synthesis model that allows
for arbitrary variation in the IMF, stellar age (for ages $>3$ Gyr),
and the detailed abundance patterns of the stars \citep{Conroy12a}.
In \citet{vanDokkum10} we used a preliminary version of this new model
to obtain constraints on the IMF in eight massive early-type galaxies
in the Virgo and Coma clusters, finding evidence for an IMF in these
galaxies that was much more bottom-heavy than the canonical Milky Way
(MW) IMF \citep{Kroupa01, Chabrier03}.  Following this work, in
\citet{vanDokkum11} we demonstrated that globular clusters in M31 with
abundance patterns similar to the massive galaxies did not show
evidence in their spectra for a bottom-heavy IMF, which was a critical
test of the technique because such clusters are known not to have
heavy IMFs from dynamical constraints \citep{Strader11}.  An
increasingly bottom-heavy IMF with increasing galaxy mass has also
been suggested by \citet{Spiniello12} based on our models and data
from the Sloan Digital Sky Survey.

Constraints on the IMF in early-type galaxies from kinematics and
gravitational lensing also favor IMFs that become increasingly
bottom-heavy toward higher galaxy masses \citep{Grillo08, Grillo10,
  Treu10, Auger10, Spiniello12, ThomasJ11, Sonnenfeld12,
  Cappellari12}.  These techniques constrain the IMF by assuming that
the stars and dark matter do not trace each other
perfectly.\footnote{The dynamical constraints imply that the
  mass-to-light ratios are larger than expected for a MW IMF, which
  can be explained either by a bottom-heavy (dwarf-dominated) IMF or a
  bottom-light (remnant-dominated) IMF.}  This assumption becomes more
plausible as the spatial distribution of the stars becomes less
spherical \citep{Cappellari12}.

Finally, constraints on the IMF from scaling relations and global
models of galaxies and dark matter also favor an IMF that becomes
increasingly bottom-heavy for more massive galaxies \citep{Dutton11b,
  Dutton12a, Dutton12b}.

This paper is the second in a series of papers aimed at measuring the
IMF in early-type galaxies.  In van Dokkum \& Conroy (2012, hereafter
Paper I) we present the sample, discuss data reduction techniques, and
explore empirical trends in the absorption line spectra.  In this
paper we apply our stellar population synthesis model to these data in
order to simultaneously measure the stellar IMF, detailed abundance
pattern, and mean stellar ages on a galaxy-by-galaxy basis.

We proceed as follows.  In Section \ref{s:data} we briefly summarize
the observations, and in Section \ref{s:model} we provide an overview
of the model.  The results are presented in Section \ref{s:res},
followed by a series of tests in Section \ref{s:tests}, a discussion
in Section \ref{s:disc} and a summary in Section \ref{s:sum}.

\section{Data}
\label{s:data}

Nearly all of the data analyzed in this paper have been obtained with
the Low Resolution Imaging and Spectrometer (LRIS) on the Keck I
telescope over the past three years.  The early-type galaxy sample
consists of a stacked spectrum of four massive galaxies in the Virgo
cluster (originally from van Dokkum \& Conroy 2010), 34 galaxies drawn
from the SAURON sample of nearby early-type galaxies \citep{Bacon01,
  deZeeuw02}, and the nuclear bulge of M31.  For all galaxies except
for the stacked Virgo spectrum and M31 the spectra are extracted
within $1/8$ of the effective radius, $R_e$.  Details of the sample
selection and data reduction can be found in Paper I.

We will also analyze spectra of four massive metal-rich globular
clusters from M31.  The red spectra for these objects were obtained
with LRIS by us \citep{vanDokkum11}, while the blue spectra were
obtained with Hectospec by Nelson Caldwell, and kindly provided to us.

\section{Model}
\label{s:model}

We employ our new stellar population synthesis (SPS) model developed
in \citet[][CvD12]{Conroy12a}, with several minor extensions.  The
core of the model is based on two empirical stellar libraries: MILES
\citep{Sanchez-Blazquez06} in the optical and IRTF \citep{Cushing05,
  Rayner09} in the near-IR, at solar metallicity.  Isochrones from the
Dartmouth \citep{Dotter08b}, Padova \citep{Marigo08}, and Lyon
\citep{Chabrier97, Baraffe98} groups are combined to provide
high-quality isochrones over the full stellar mass range appropriate
for older stellar populations.  SPS models are then constructed for
ages ranging from $3-13.5$ Gyr and for arbitrary IMFs by combining the
isochrones with the empirical libraries.  Constructing models with
younger ages was not possible because the IRTF library does not
currently include hot stars.  In CvD12 we also computed a grid of
fully synthetic stellar spectral libraries in order to model the
variation of individual elemental abundances.  In its present version,
the model allows for variation in the elements C, N, Na, Mg, Si, Ca,
Ti, Cr, Mn, and Fe, and O,Ne,S are varied in lock-step.  We emphasize
that the synthetic spectra are only used differentially with respect
to the empirical libraries\footnote{The synthetic spectra are in fact
  used in an absolute sense to bridge the gap in wavelength coverage
  between the MILES and IRTF libraries at $0.74\mu m-0.8\mu m$.  This
  wavelength range is not used in the present work.}.  The model makes
predictions over the full optical and near-IR wavelength range of
$0.36\mu m<\lambda<2.4\mu m$.

All models that feature simultaneous variation in multiple elements
are constructed by assuming that the effect of each element on the
spectrum is independent of the others.  For example, a model with
[Mg/Fe]=+0.2 and [N/Fe]=+0.2 is created by combining two models, one
with only Mg variation and one with only N variation.  This is an
assumption in every model that includes variation in multiple
abundance patterns \citep[e.g.,][]{Thomas03, Schiavon07}.  To our
knowledge, this assumption has not been rigorously tested.  We simply
note it here as a possible systematic uncertainty in the modeling.
This issue will be addressed in future work.

Moreover, we make the standard assumption that the response of the
spectrum to variation in an element is linear in [e/Fe], where `e'
stands for a generic element.  In reality, the variation of spectral
features with [e/Fe] is not linear in [e/Fe] when the range in [e/Fe]
is large, especially for strong spectral features and for elements
that play an important role in the atmospheric structure of stars.
For most purposes this non-linearity is not important, as the derived
abundances do not vary significantly beyond the abundance range used
to create the models (for example, our model spectra were computed
with variations of $\pm0.3$ dex for most elements).  As we will see in
later sections, this is not true for the derived [Na/Fe] abundances,
which reach as high as 1 dex in our best-fit models.  We have
therefore computed new stellar spectral models with [Na/Fe] variation
up to 0.9 dex and we use these new response functions in the present
model.  The use of these models, rather than linear extrapolations
from the 0.3 dex models, results in a derived [Na/Fe] that is lower by
$\approx0.2$ dex for the most extreme cases.  Moreover, there is some
evidence that non-LTE effects will result in stronger NaI lines
compared to LTE models \citep{Lind11}.  To first order these effects
should only change the derived [Na/Fe] abundance, but further work
will be required to assess whether or not non-LTE effects can impact
the derived IMF values as well.

Notice that since abundance ratio variations are grafted onto
empirical stellar spectra at approximately solar metallicity, the
model is only applicable to systems with metallicities not too
different from solar.  The synthetic spectra allow us to create models
with non-solar metallicities, but the extrapolation becomes less
reliable as the metallicity deviates significantly from solar.

The model also allows for variation in the effective temperature,
$\teff$, of the individual stars.  This was implemented by computing
new synthetic spectra with different $\teff$ and then differentially
modifying the empirical spectra.  This parameter allows us to explore
the effect of variation in the isochrones with metallicity.  For
example, a change in [$\alpha$/Fe] by $\pm0.2$ or [Fe/H] by $\pm0.1$
dex results in a roughly 50K change in the location of the isochrones
near solar metallicity \citep[e.g.,][]{Dotter07, Dotter08b}.  Most SPS
models that consider variable abundance patterns do not include the
abundance effects on the isochrones \citep[e.g.,][]{Thomas03,
  Graves08, Thomas11}, with the exception of the models of
\citet{Coelho07} and \citet{LeeHC09}.

We emphasize that these models are the only ones currently available
that follow the effect of individual elemental abundance variation on
the full spectrum.  To our knowledge, all other models that include
variation in individual elemental abundances do so only on the effect
of spectral indices, principally the Lick index system
\citep[e.g.,][]{Thomas03, Schiavon05, LeeHC09}.

In CvD12 we considered IMFs that were a single power-law over the full
mass range.  In the present work we consider a three component
power-law with separate indices, $\alpha_1$, $\alpha_2$, and
$\alpha_3$, describing respectively the $0.1\Msun<M<0.5\Msun$,
$0.5\Msun<M<1.0\Msun$, and $1.0\Msun<M<100\Msun$ mass intervals.  For
reference, a Salpeter IMF has $\alpha_1=\alpha_2=\alpha_3=2.3$ and a
\citet{Kroupa01} IMF has $\alpha_1=1.3, \alpha_2=2.3$, $\alpha_3=2.3$.
Here we will fit for $\alpha_1$ and $\alpha_2$ and will fix
$\alpha_3=2.3$.  Stellar remnants are included in our $M/L$ values
according to a standard prescription \citep{Conroy09a}.  A
\citet{Kroupa01} IMF is adopted as our reference `MW' IMF.

We have also added several additional parameters meant to capture the
addition of minority stellar populations.  In addition to the age of
the bulk population, we include a parameter to allow `frosting'
\citep{Trager00} with a young population with an age of 3 Gyr (the
youngest age in our model).  We also include two parameters describing
the contribution to the light from hot stars (whether they be young
stars or hot horizontal branch stars).  One of these parameters is the
temperature of the star (ranging from $1-3\times10^4$ K) and the
second is the fraction of the total flux comprised of these stars.
Finally, we allow for the addition of arbitrary amounts of M giant
light (specifically an M7 giant).  For our purposes we consider these
as nuisance parameters, as our main goal is to measure the IMF from
integrated light.

The fiducial model is characterized by 19 parameters: 11 for the
abundance pattern, one for the age, one for $\teff$ offsets, two for
the IMF, two for the hot star component, one for the young component,
and one for the fraction of additional M giant light.  In addition to
these 19 parameters, we also simultaneously fit for the velocity
dispersion and redshift of each galaxy, bringing the total number of
parameters to be fit to 21.  These parameters are summarized in Table
\ref{t:params}.

The native resolution of the model is not constant in velocity.  In
order to facilitate comparison to the data, we have broadened the
models to a constant velocity dispersion of $100\kms$.  We note that
the observed spectra have also been convolved to a resolution that is
constant in velocity, with a dispersion of $100\kms$ (see Paper I for
details).

\begin{figure*}[!t]
\center
\resizebox{7in}{!}{\includegraphics{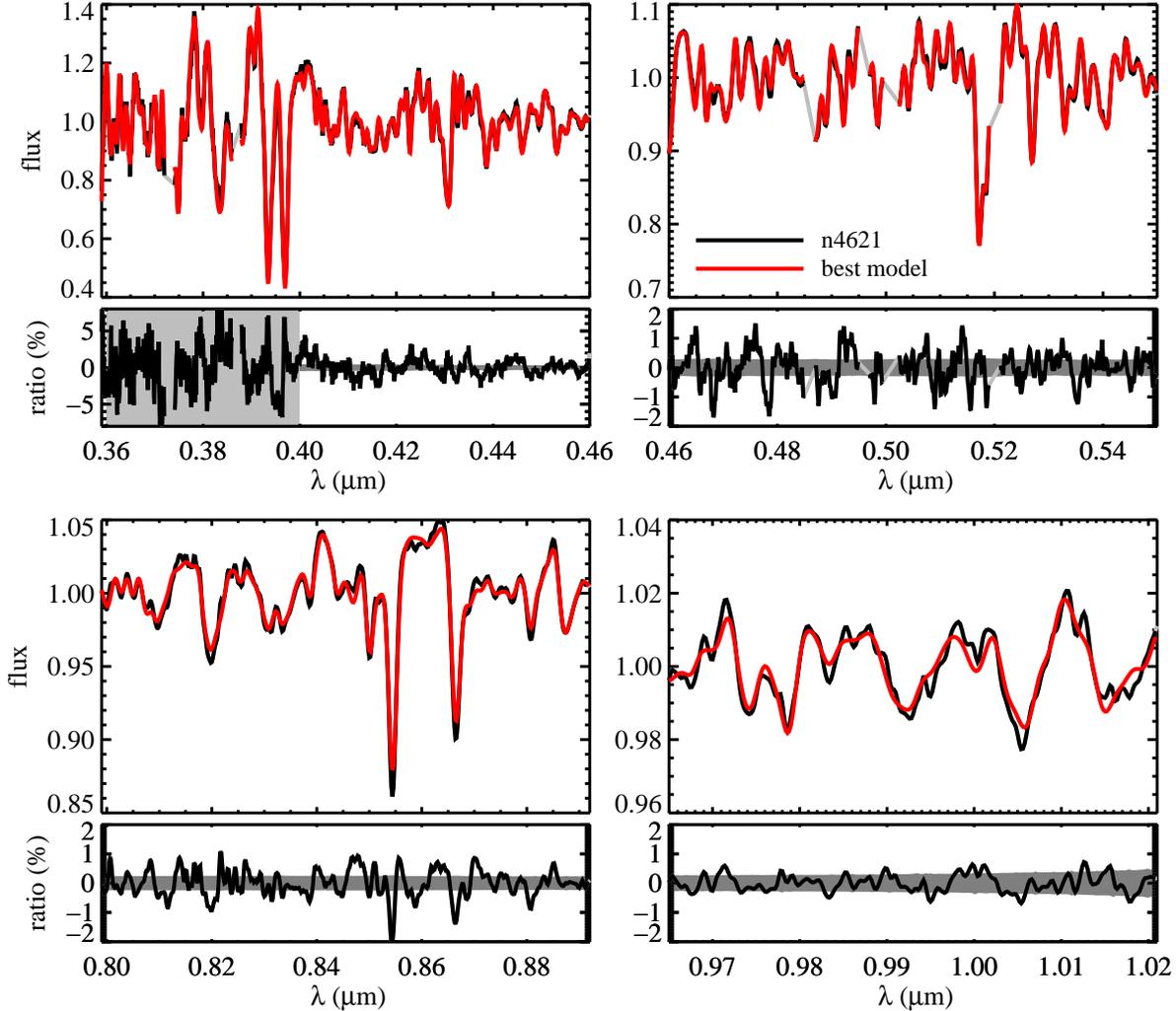}}
\caption{Comparison between the spectrum of NGC 4621 and the best-fit
  model.  Within each plotted wavelength interval,
  continuum-normalized fluxes are shown in the top panel and the ratio
  between model and data are shown in the bottom panel.  The grey
  shaded bands demarcate the noise limits of the data.  The data at
  $\lambda<0.4\mu m$ are not used in the fit; the grey shaded bands
  become very large in that wavelength range to highlight this fact.
  The light grey spectral regions are masked from the fit because of
  possible emission line contamination.  NGC 4621 has one of the most
  bottom-heavy IMFs in our sample.}
\label{fig:bestspec}
\end{figure*}

\begin{figure*}[!t]
\center
\resizebox{7in}{!}{\includegraphics{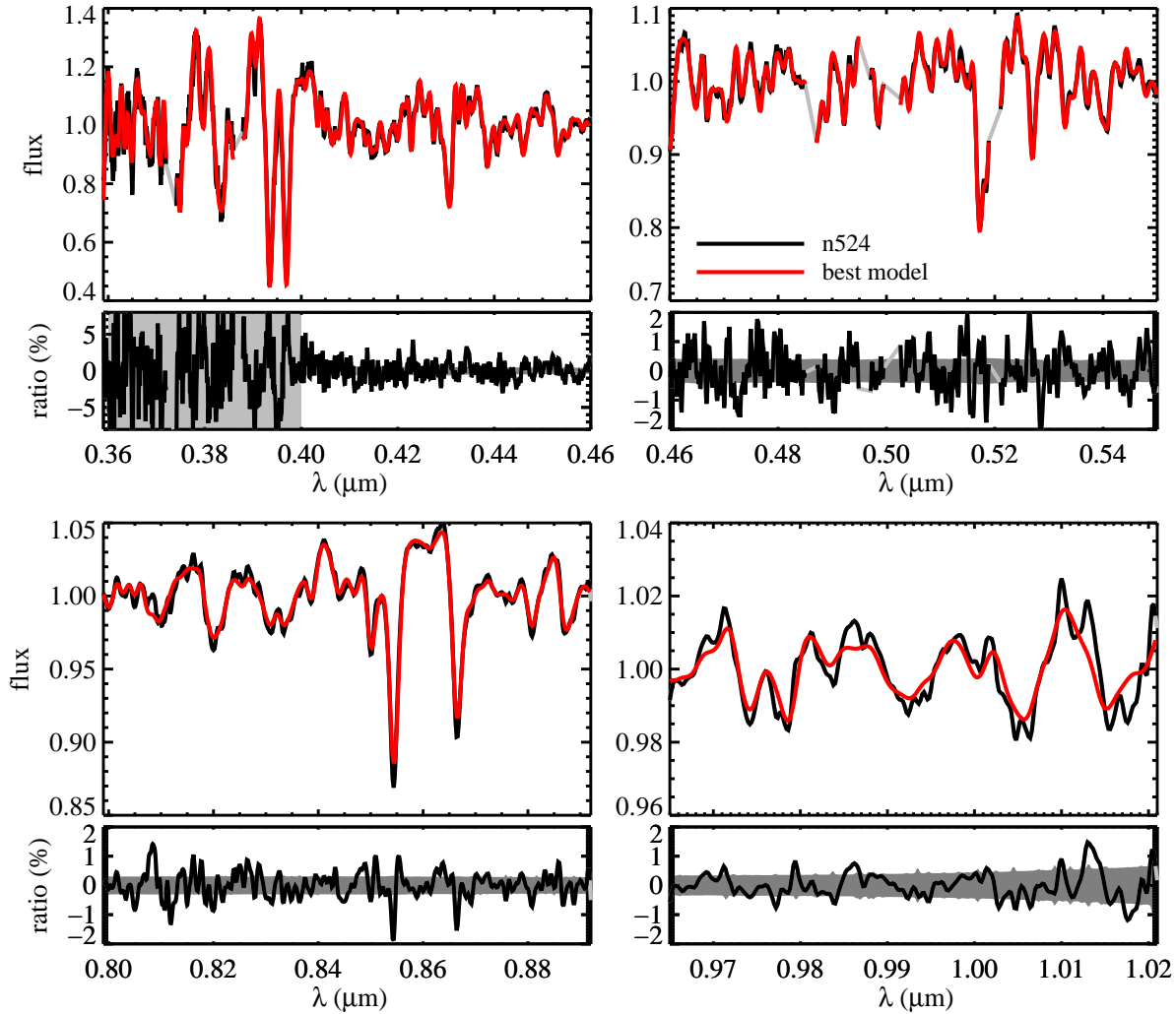}}
\caption{Comparison between the spectrum of NGC 524 and the best-fit
  model; see Figure \ref{fig:bestspec} for details.  NGC 524 has an
  inferred IMF similar to the MW.}
\label{fig:bestspec2}
\end{figure*}

\begin{figure*}[!t]
\center
\resizebox{7.6in}{!}{\includegraphics{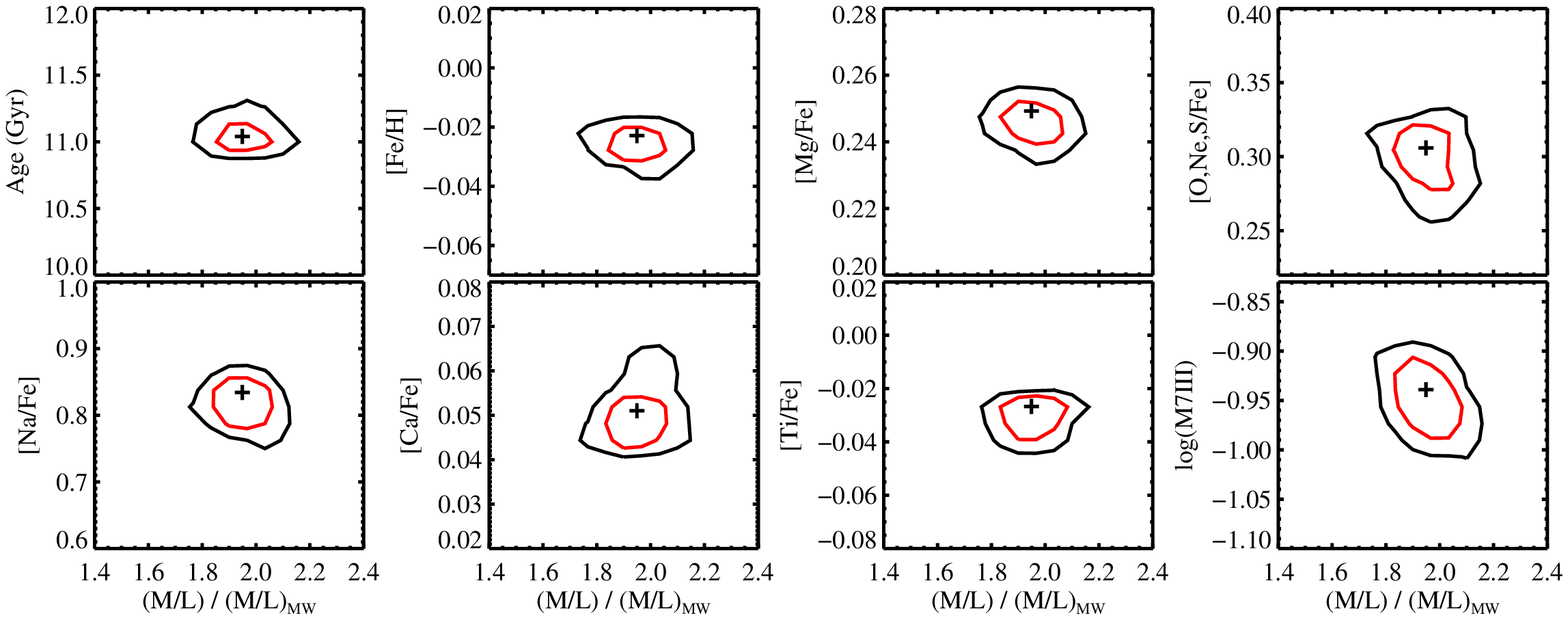}}
\vspace{0.1cm}
\caption{Covariance between various parameters and the normalization
  of the IMF.  The latter is displayed as a ratio between the
  measured $M/L$ and the $M/L$ assuming a MW IMF.  Results are shown
  for the fit to NGC 4621.  The contours represent the 68\% and 95\%
  confidence limits (red and black lines, respectively).  The symbols
  mark the median of the marginalized likelihoods.  Notice that there
  is weak or no correlation between the various parameters and the
  IMF, and that each parameter is very well constrained.}
\label{fig:mcmc}
\end{figure*}

\subsection{Fitting Procedure}
\label{s:fit}

We adopt a Markov Chain Monte Carlo (MCMC) fitting technique in order
to efficiently explore the large parameter space of the model.  In the
MCMC algorithm, a step is taken in parameter space and this step is
accepted if the new location has a lower $\chi^2$ compared to the
previous location, and is accepted with probability $e^{-\Delta
  \chi^2/2}$ if the $\chi^2$ is higher than the previous location.
Each step in the chain is recorded.  After a sufficient number of
steps the likelihood surface produced by the chain will converge to
the true underlying likelihood.  Convergence is defined according to
the prescription described in \citet{Dunkley05}.  We have found that
$\sim10^5$ steps are required in order to achieve convergence in all
parameters.  The `burn-in' region, where the chain is descending to
the minimum of $\chi^2$, is removed before analyzing the chain, as is
standard practice.

\begin{figure}[!t]
\center
\resizebox{3.5in}{!}{\includegraphics{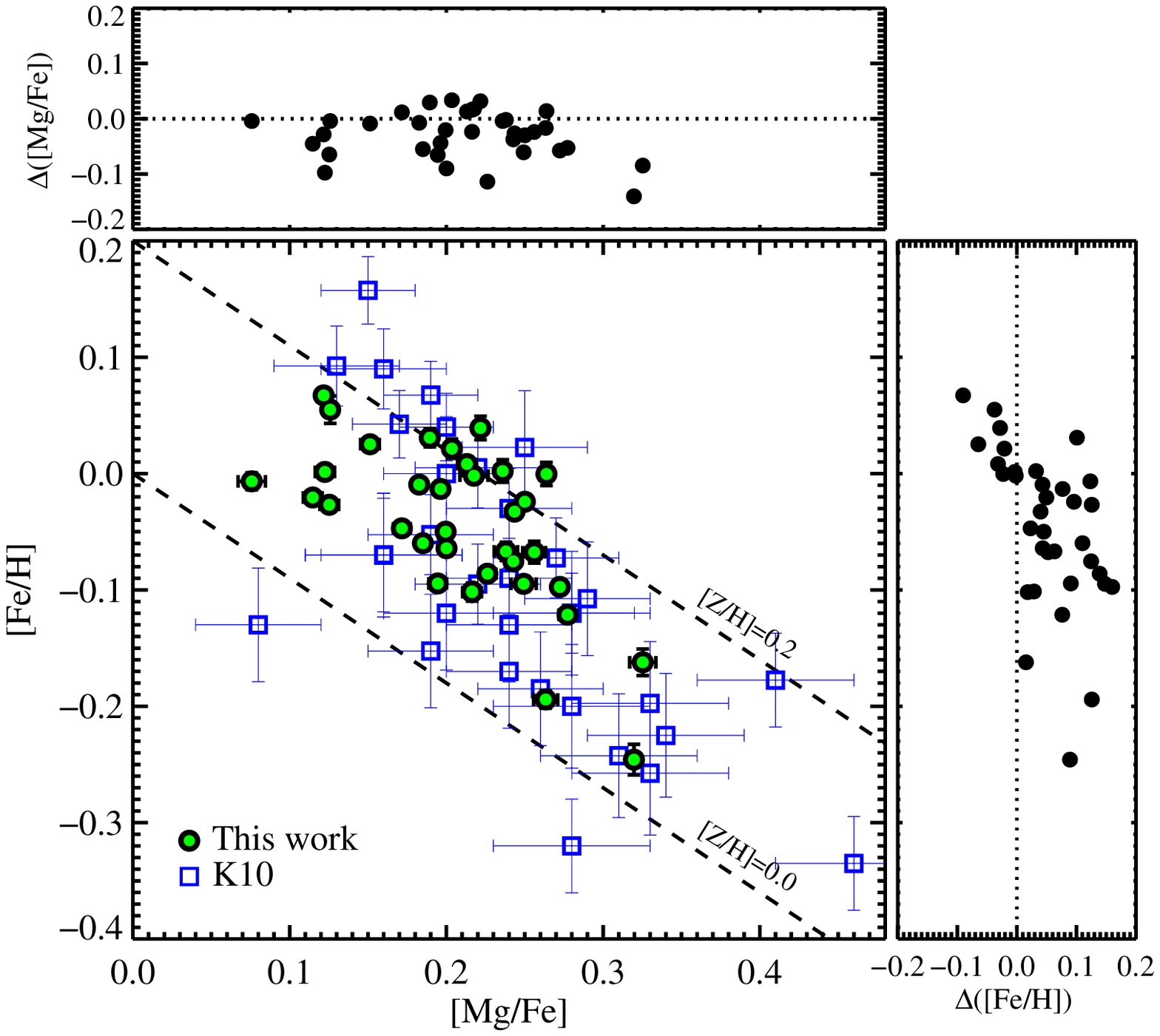}}
\caption{Comparison between [Fe/H] and [Mg/Fe] values derived in this
  work and in K10.  The formal statistical errors on our measurements
  are often smaller than the symbol size.  Dashed lines represent
  constant total metallicity, [Z/H].  The upper and left panels show
  differences between our and K10's measurements as a function of our
  values.}
\label{fig:k10}
\end{figure}

\begin{figure*}[!t]
\center
\resizebox{7in}{!}{\includegraphics{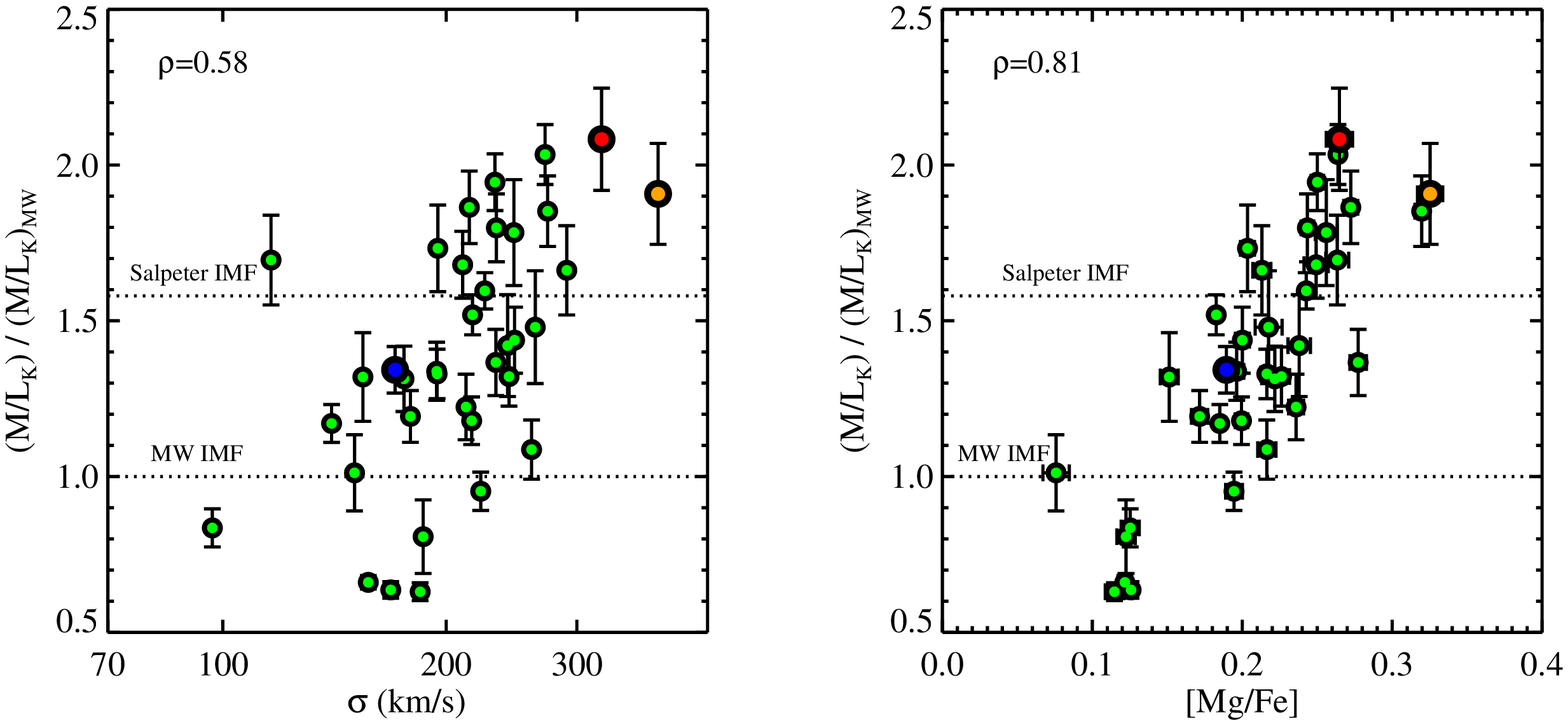}}
\caption{Best-fit $\mlk$ in units of the best-fit $\mlk$ for a MW IMF.
  This quantity is sensitive only to the IMF.  Results are shown as a
  function of $\sigma$ (left panel) and [Mg/Fe] (right panel).  All
  galaxies would lie on the dotted line if the IMF was universal and
  of the MW form.  The red symbol is the result for the stacked
  spectrum of four massive early-type galaxies from the Virgo cluster
  studied in \citet{vanDokkum10}, the orange symbol is M87, and the
  blue symbol is M31.  The Spearman correlation coefficient, $\rho$,
  is included in the legend of each panel.  The data therefore support
  a scenario wherein the IMF becomes steeper with increasing $\sigma$
  and/or [Mg/Fe].}
\label{fig:imf1}
\end{figure*}

Priors on the model parameters must be specified in any MCMC algorithm
to keep the chain from wandering into unphysical or unrealistic
regions of parameter space.  Our choice of priors on each parameter
are specified in Table \ref{t:params}.  Outside of the prior range,
$\chi^2$ is penalized by a Gaussian with a width of $\sigma_p=0.01$.
The adopted priors have little impact on our results because the
parameters are always well-constrained within the prior range.

In the present work we are most interested in the information
contained in narrow spectral features.  Therefore, we compute $\chi^2$
after dividing the model and data spectrum by a high-order polynomial.
Operationally, we split the spectrum into four wavelength intervals:
$0.4\mu m-0.46\mu m$, $0.46\mu m-0.55\mu m$, $0.80\mu m-0.89\mu m$,
$0.96\mu m-1.02\mu m$, and divide each sub-region by a polynomial of
degree $n$ where $n\equiv (\lambda_{\rm max}-\lambda_{\rm
  min})/100$\AA. Variation in $n$ by $\pm2$ induces changes to the
spectrum at the $\lesssim0.1\%$ level.

The wavelength range $0.89-0.96\mu m$ is not included because the sky
absorption corrections are significant in that wavelength interval
(see Paper I for details).  While both the models and data extend to
$\lambda<0.4\mu m$, the CaII H\&K lines at $0.39\mu m$ are not
well-modeled at the sub-percent level because these features form at
very low densities in the stellar atmosphere.  They are thus subject
to non-LTE and 3D effects, which are not included in the calculation
of our synthetic spectra.  As our goal is to model the observed
spectra at the sub-percent level, we ignore the very blue end of the
spectra.  Nonetheless, the best-fit models, constrained at
$\lambda>0.4\mu m$, do provide a good fit to the data at
$\lambda<0.4\mu m$, with typical residuals at the several percent
level.  Examples of the quality of fit in the very blue are given in
Figures \ref{fig:bestspec} and \ref{fig:bestspec2}, discussed below.

We emphasize that the information contained in the broadband shape of
the spectral energy distribution is not used in the fitting procedure.
We choose this approach because it is very difficult to achieve
accurate flux calibration over a large wavelength baseline at the
percent level.  

We mask the spectral regions surrounding the H$\beta$, [OIII] and [NI]
emission lines and any spectral region contaminated by cosmic rays.

\section{Constraints on the IMF}
\label{s:res}

In this section we present constraints on the stellar IMF for a sample
of early-type galaxies, the nuclear bulge of M31, and globular
clusters.  Other parameters derived from our stellar population
synthesis model, including detailed abundance patterns, will be
presented in future work.

\begin{deluxetable*}{rcccccccccc}
\tablecaption{Results from Stellar Population Synthesis Modeling}
\tablehead{ \colhead{Galaxy} & \colhead{$\langle$S/N$\rangle$} & \colhead{rms ($\%$)} & 
\colhead{$\chi^2/$dof}&\colhead{$\sigma$ (km s$^{-1}$)}&\colhead{[Mg/Fe]}&\colhead{[Fe/H]}&
\colhead{$M/L_r$}&\colhead{$M/L_I$} & 
\colhead{$M/L_K$} &\colhead{$(M/L_K)_{\rm MW}$} }
\startdata
        M 31 &   439 &   0.86 &  10.21 &  170 &   0.19 &   0.03 &   6.17 &   3.69 &   1.20 &   0.89 \\
     NGC 474 &   219 &   0.76 &   2.13 &  178 &   0.17 &  -0.05 &   2.98 &   1.88 &   0.69 &   0.58 \\
     NGC 524 &   259 &   0.82 &   3.11 &  260 &   0.22 &  -0.10 &   3.80 &   2.36 &   0.85 &   0.78 \\
     NGC 821 &   266 &   0.69 &   2.55 &  216 &   0.20 &  -0.05 &   3.71 &   2.30 &   0.81 &   0.68 \\
    NGC 1023 &   440 &   0.63 &   6.14 &  217 &   0.18 &  -0.01 &   5.26 &   3.20 &   1.04 &   0.68 \\
    NGC 2549 &   371 &   0.76 &   6.28 &  157 &   0.12 &   0.07 &   1.92 &   1.22 &   0.42 &   0.63 \\
    NGC 2685 &   189 &   0.95 &   2.27 &   96 &   0.13 &  -0.03 &   1.80 &   1.16 &   0.41 &   0.50 \\
    NGC 2695 &   221 &   0.75 &   2.11 &  210 &   0.25 &  -0.09 &   5.81 &   3.56 &   1.28 &   0.76 \\
    NGC 2699 &   180 &   0.81 &   1.72 &  154 &   0.15 &   0.03 &   3.80 &   2.39 &   0.87 &   0.66 \\
    NGC 2768 &   239 &   0.89 &   2.97 &  222 &   0.19 &  -0.09 &   2.98 &   1.85 &   0.64 &   0.67 \\
    NGC 2974 &   260 &   0.75 &   2.90 &  247 &   0.20 &  -0.06 &   5.22 &   3.21 &   1.16 &   0.80 \\
    NGC 3377 &   352 &   0.77 &   5.88 &  140 &   0.19 &  -0.06 &   2.52 &   1.60 &   0.57 &   0.49 \\
    NGC 3379 &   431 &   0.65 &   5.96 &  225 &   0.24 &  -0.08 &   5.58 &   3.40 &   1.12 &   0.70 \\
    NGC 3384 &   432 &   0.71 &   7.74 &  168 &   0.13 &   0.05 &   2.08 &   1.30 &   0.41 &   0.65 \\
    NGC 3414 &   264 &   0.84 &   3.89 &  243 &   0.23 &  -0.09 &   4.92 &   3.00 &   1.04 &   0.79 \\
    NGC 3608 &   271 &   0.76 &   3.25 &  194 &   0.22 &  -0.10 &   4.31 &   2.67 &   0.93 &   0.70 \\
    NGC 4262 &   227 &   0.68 &   1.78 &  214 &   0.27 &  -0.10 &   6.96 &   4.21 &   1.45 &   0.78 \\
    NGC 4270 &   132 &   0.80 &   0.88 &  150 &   0.08 &  -0.01 &   2.38 &   1.54 &   0.60 &   0.59 \\
    NGC 4278 &   286 &   1.05 &   6.23 &  273 &   0.32 &  -0.25 &   8.68 &   5.08 &   1.61 &   0.87 \\
    NGC 4382 &   328 &   0.66 &   4.18 &  184 &   0.11 &  -0.02 &   1.05 &   0.69 &   0.25 &   0.40 \\
    NGC 4458 &   167 &   0.97 &   2.08 &  116 &   0.26 &  -0.19 &   3.75 &   2.37 &   0.94 &   0.56 \\
    NGC 4459 &   269 &   0.71 &   2.70 &  186 &   0.12 &   0.00 &   1.94 &   1.26 &   0.46 &   0.57 \\
    NGC 4473 &   270 &   0.58 &   2.03 &  194 &   0.20 &  -0.01 &   4.56 &   2.82 &   1.01 &   0.75 \\
    NGC 4486 &   244 &   0.82 &   3.17 &  385 &   0.33 &  -0.16 &   9.82 &   5.66 &   1.71 &   0.90 \\
    NGC 4546 &   286 &   0.73 &   3.04 &  233 &   0.24 &  -0.03 &   7.16 &   4.30 &   1.40 &   0.78 \\
    NGC 4552 &   329 &   0.68 &   3.90 &  271 &   0.26 &  -0.00 &   8.41 &   4.97 &   1.55 &   0.76 \\
    NGC 4564 &   238 &   0.76 &   2.52 &  175 &   0.22 &   0.04 &   5.14 &   3.10 &   0.96 &   0.73 \\
    NGC 4570 &   229 &   0.67 &   1.82 &  194 &   0.20 &   0.02 &   7.19 &   4.27 &   1.37 &   0.79 \\
    NGC 4621 &   343 &   0.68 &   4.30 &  232 &   0.25 &  -0.02 &   7.20 &   4.34 &   1.39 &   0.71 \\
    NGC 4660 &   240 &   0.64 &   1.79 &  212 &   0.24 &   0.00 &   4.50 &   2.74 &   0.89 &   0.73 \\
    NGC 5308 &   167 &   0.60 &   0.82 &  241 &   0.24 &  -0.07 &   4.76 &   2.93 &   1.03 &   0.73 \\
    NGC 5813 &   204 &   0.86 &   2.33 &  233 &   0.28 &  -0.12 &   4.62 &   2.85 &   1.04 &   0.76 \\
    NGC 5838 &   231 &   0.62 &   1.46 &  290 &   0.21 &   0.01 &   5.86 &   3.57 &   1.21 &   0.73 \\
    NGC 5845 &   166 &   0.57 &   0.70 &  263 &   0.22 &  -0.00 &   5.38 &   3.29 &   1.09 &   0.74 \\
    NGC 5846 &   192 &   0.87 &   1.95 &  246 &   0.26 &  -0.07 &   7.23 &   4.34 &   1.46 &   0.82 
\enddata
\vspace{0.1cm} 

\tablecomments{The typical fractional error on $M/L$ is 7\%
  (statistical).  Tests performed in Section \ref{s:sys} suggest that
  the systematic errors on $M/L$ are not larger than 50\%.  The
  typical fractional error on $(M/L)_{\rm MW}$ is 2\% (statistical).
  The formal errors on [Mg/Fe] and [Fe/H] are $<1$\%.  All quantities
  are measured within an effective circular aperture of radius $R_e/8$
  except for M31, which is within the central $4"=15$ pc.}
\label{t:res}
\end{deluxetable*}

\subsection{Early-Type Galaxies}

A typical fit to the continuum normalized spectrum of an early-type
galaxy is shown in Figures \ref{fig:bestspec} and \ref{fig:bestspec2}
for NGC 4621 and NGC 524, respectively.  In these figures we compare
the observed spectra to the best-fit models.  We also plot the ratio
between the model and data and compare to the noise-to-signal ratio of
the data.  In an Appendix we show the fits to the other 33 galaxies
from Paper I.

In Table \ref{t:res} we provide the mean S/N of the spectrum (averaged
over the wavelength range used in the fit), the rms residual between
the best-fit model and data, the minimum $\chi^2$/dof, the best-fit
velocity dispersion, [Mg/Fe], and [Fe/H], the best-fit mass-to-light
ratios in the $r$, $I$ and $K-$bands allowing for IMF variation, and
the best-fit mass-to-light ratio assuming a fixed MW \citep{Kroupa01}
IMF.  All quantities are measured within an effective circular
aperture of radius $R_e/8$ except for M31, which is within the central
$4"=15$ pc.

In Figure \ref{fig:mcmc} we show the covariance between several
derived parameters and the mass-to-light ratio, $M/L$, for NGC 4621.
The latter quantity is shown in units of the mass-to-light ratio
assuming a MW IMF, $(M/L)_{\rm MW}$.  Constraints on the IMF will be
plotted in this way throughout the paper in order to isolate the
effects of the IMF on $M/L$ from the effects of age, [Z/H], etc.  The
most important point to take from this figure is the very weak or
absence of a correlation between the IMF and other parameters.  We
have inspected a much greater cross-section of parameter space than
what is shown in Figure \ref{fig:mcmc} and find in general very little
correlation between parameters, indicating that each parameter is
well-constrained by the data. It is also evident from this figure that
the derived $M/L$ is much larger than expected for a MW IMF.  Indeed,
NGC 4621 has one of the steepest IMFs in our sample (a Salpeter IMF is
$\approx60\%$ heavier than a MW IMF, and so NGC 4621, with an $M/L$
that is twice as heavy as a MW IMF, has an IMF even steeper than
Salpeter).

NGC 4621 also has one of the highest [Na/Fe] values in our sample.
The only sodium line being fit is the NaI doublet at $0.82\mu m$; the
NaD feature at $0.59\mu m$ is not covered in the observed spectra.
The constraint on [Na/Fe] is therefore coming from a combination of
the feature at $0.82\mu m$ and the fact that the sodium abundance has
an indirect effect on the entire spectrum due to its influence on the
free electron abundance in stellar atmospheres (see CvD12 for
details).  It is important to notice that even when allowing the
[Na/Fe] abundance to reach very high values, a bottom-heavy IMF is
still favored for this particular galaxy.  In other words, the very
strong NaI feature is due {\it both} to high [Na/Fe] abundance and a
preponderance of low-mass stars.  A similar result was found by
\citet{Spiniello12} who analyzed the spectra of two higher-redshift
early-type galaxies.  Spectra that cover the NaD feature will be very
valuable for placing upper limits on the [Na/Fe] abundance.  Of
course, the NaD feature can only place an upper limit on the [Na/Fe]
abundance because this feature may also be influenced by absorption
from gas in the interstellar medium.

Before considering constraints on the IMF for the full sample, we
first compare our best-fit [Fe/H] and [Mg/Fe] abundances to the [Fe/H]
and [$\alpha$/Fe] abundances \footnote{K10 quote total metallicity, Z,
  rather than [Fe/H].  We have converted their results to [Fe/H] via
  the equation [Fe/H]$=$Z$-0.75$[$\alpha$/Fe].  This relation was
  derived from tabulated values of [Fe/H], Z, and [$\alpha$/Fe] kindly
  provided to us by R. Schiavon.} derived for the same galaxies by
\citet[][K10]{Kuntschner10}.  The latter are based on the SPS model of
\citet{Schiavon07}.  This comparison is shown in Figure \ref{fig:k10}.
In K10 all of the $\alpha$ elements track each other (except for Ti),
so [Mg/Fe]$=$[$\alpha$/Fe] in that work. Notice that the comparison is
only approximate because K10 derive abundances within true circular
apertures with radii equal to $R_e/8$, while we derive abundances
within a slit of radius $R_e/8$ with a radial weighting meant to mimic
a circular aperture (see Paper I).  If the galaxies were perfectly
smooth in the azimuthal direction, then the two approaches would yield
identical spectra.  However, it is clear from the 2D stellar
population maps in K10 that there is significant variation in the
stellar populations at fixed radius.  In any event, the iron
abundances and $\alpha-$enhancements generally agree to within
$\lesssim0.1$ dex (the scatter between the two methods is 0.05 and
0.07 for [Mg/Fe] and [Fe/H], respectively), which is encouraging
considering the different apertures and modeling techniques.  It is
also evident from Figure \ref{fig:k10} that there is an
anti-correlation between [Fe/H] and [Mg/Fe] such that the total range
in metallicity, [Z/H], is only $\approx0.2$ dex.

The principal result of this paper is shown in Figure \ref{fig:imf1}.
In this figure we show the $K-$band mass-to-light ratio, $\mlk$
normalized to the mass-to-light ratio expected for a MW IMF.  This
quantity is directly related to the IMF and is plotted as a function
of galaxy velocity dispersion, $\sigma$, and [Mg/Fe].  Error bars are
marginalized 68\% confidence limits.  The most important conclusion to
draw from this figure is that significant galaxy-to-galaxy variation
in the IMF is inferred, ranging from IMFs slightly lighter than the
MW, to IMFs significantly heavier than the MW.  In these units a
Salpeter IMF has a value of $\approx1.6$, so there are galaxies that
have inferred IMFs steeper even than Salpeter.

\begin{figure}[!t]
\center
\resizebox{3.5in}{!}{\includegraphics{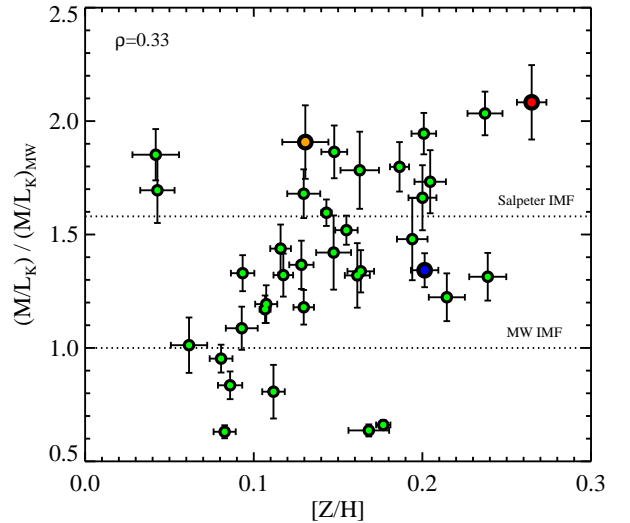}}
\caption{Best-fit IMF (plotted as in Figure \ref{fig:imf1}) as a
  function of the best-fit total metallicity, [Z/H]. The Spearman
  correlation coefficient, $\rho$, is included in the legend.  The
  correlation between IMF and [Z/H] is much weaker than between either
  $\sigma$ or [Mg/Fe], suggesting that total metallicity is not the
  key variable driving IMF variation.  This plot also implies that any
  model uncertainties that correlate with [Z/H] are unlikely to affect
  the qualitive IMF variation displayed in Figure \ref{fig:imf1}.}
\label{fig:imfz}
\end{figure}

In Figure \ref{fig:imf1} we also see clear evidence for correlations
between the IMF and $\sigma$ and between the IMF and [Mg/Fe].  The
Spearman rank correlation coefficients, $\rho$, are 0.58 and 0.81, for
the IMF correlation with $\sigma$ and [Mg/Fe], respectively.  [Mg/Fe]
is an indicator of the star formation timescale, and the fact that the
IMF appears to correlate more strongly with this quantity than with
$\sigma$ suggests that the intensity of star formation on kpc scales
may play a key role in determining the shape of the IMF \citep[see
also][who reach similar conclusions based on analyzing stacked spectra
of galaxies in the Coma cluster]{Smith12b}.  We return to this topic
in Section \ref{s:origin}.

Figure \ref{fig:imfz} shows the best-fit $\mlk$ ratios as a function
of the total metallicity, [Z/H], estimated as
[Z/H]$=$[Fe/H]$+0.94$[Mg/Fe] \citep{Thomas03}.  The correlation here
is much weaker than with either $\sigma$ or [Mg/Fe] ($\rho=0.33$),
suggesting that total metallicity is not the fundamental variable
driving IMF variation.  The lack of correlation between [Z/H] and IMF
is also an important diagnostic for any potential model systematic
uncertainties that may correlate with metallicity, as we discuss
further in $\S$\ref{s:imfnec}.

\subsubsection{Quality of Fit}

Our constraints on the IMF are derived by marginalizing over 19 other
parameters, including 11 parameters characterizing the detailed
abundance pattern and four nuisance parameters meant to capture
uncertain aspects of stellar population synthesis modeling.  Despite
this high degree of flexibility, the error bars for most galaxies are
inconsistent with a MW IMF.

Nonetheless, we might also ask how well-fit are the data by models in
which the IMF is fixed to the MW value for all galaxies.  We have thus
re-fit each galaxy in our sample with a model that is identical to the
fiducial model except that the IMF is fixed to the MW value
\citep[specifically a][IMF]{Kroupa01}.  Inspection of the best-fit
models confirms our expectation that the Wing-Ford band at $0.99\mu m$
is a powerful IMF diagnostic when considered in conjunction with other
spectral features \citep{Conroy12a}.  In contrast, the NaI feature at
$0.82\mu m$ can be reasonably well-fit with a model without IMF
variation but with higher [Na/Fe] abundance.  The quality of the fit
over the full spectral range for such models is, however, poorer (see
below).

\begin{figure}[!t]
\center
\resizebox{3.5in}{!}{\includegraphics{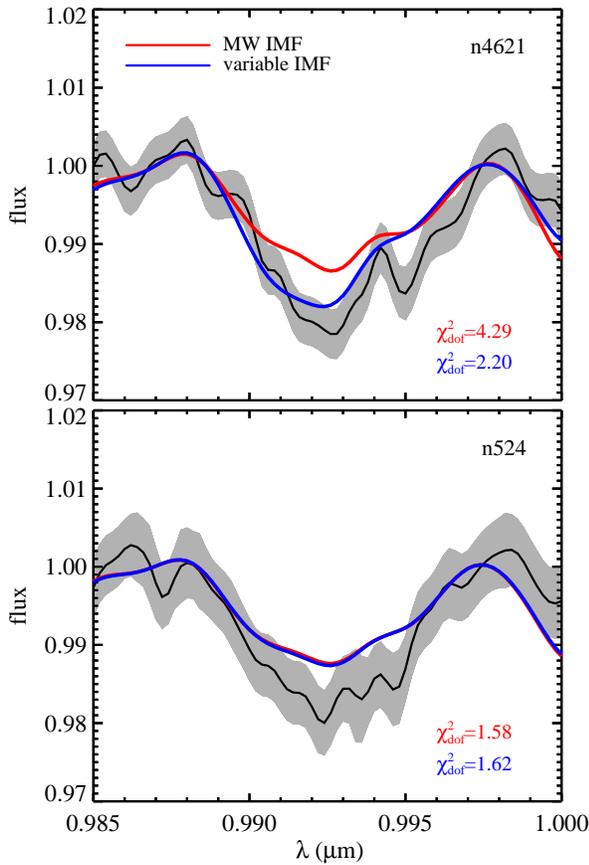}}
\caption{Zoom-in on the IMF-sensitive Wing-Ford band for NGC 4621
  (top) and NGC 524 (bottom).  These are the two galaxies shown in
  Figures 1 and 2.  The observed spectrum (black line and grey band
  encompassing the $1\sigma$ errors) is compared to several best-fit
  models. The fiducial model, which fits the full wavelength range and
  includes parameters for the IMF (blue line), is compared to a model
  where the IMF is fixed to the MW value (red line).  The $\chi^2_{\rm
    min}$ is shown for each model fit, computed over the range
  $0.988\mu m-0.997\mu m$.  The spectrum of NGC 4621 is poorly fit by
  a MW IMF, while the spectrum of NGC 524 is equally well-fit by a
  fixed or variable IMF model.  Notice that the best-fit models for
  NGC 524 are formally acceptable fits to the data around the
  Wing-Ford band, despite the appearance that the models slighly
  underpredict the feature strength.}
\label{fig:wfb}
\end{figure}

As an example of the quality of the fits, we show in Figure
\ref{fig:wfb} a zoom-in on the Wing-Ford band for the two galaxies
shown in Figures 1 and 2.  In this figure the data are compared to two
models, one in which the IMF is allowed to vary in the fit, and a
second in which the IMF is fixed to the MW value.  In the case of NGC
4621, where the best-fit IMF is significantly different from the MW
IMF, we see that a fixed IMF model fails to capture the depth of the
observed Wing-Ford band.  This suggests that the Wing-Ford feature can
be a strong discriminant between fixed and variable IMF models.  In
the case of NGC 524 there is essentially no difference between the
predictions of the fixed and variable IMF models.  This occurs because
the best-fit variable IMF model returns a MW IMF for this galaxy.
Nonethelss, both models predict a Wing-Ford band that is slightly too
weak compared to the data for this galaxy.  Formally the fits are
acceptable, and we remind the reader that the data points are highly
covariant owing to the velocity broadening of the spectrum.

\begin{figure}[!t]
\center
\resizebox{3.5in}{!}{\includegraphics{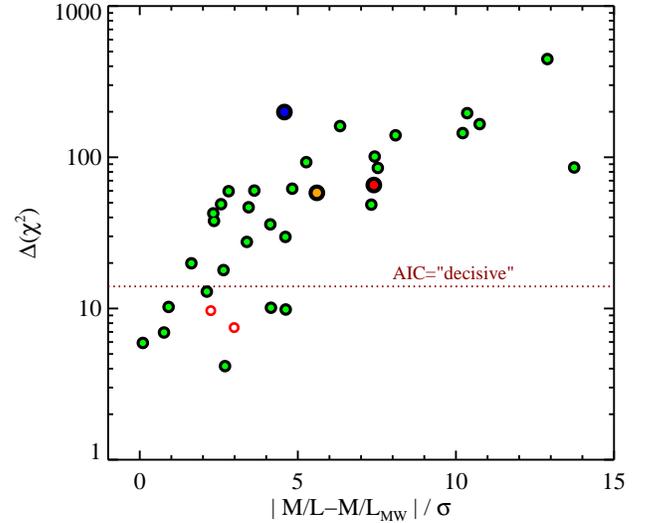}}
\caption{Difference in the minimum of $\chi^2$ between models with and
  without variation in the IMF, as a function of the difference in
  best-fit $\mlk$ in units of the uncertainty.  Symbol colors are as
  in Figure \ref{fig:imf1}.  The two open symbols have
  $\Delta(\chi^2)<0$ and so for these galaxies we plot
  $|\Delta(\chi^2)|$.  The model with IMF variation has two additional
  degrees of freedom compared to the fixed IMF model.  In the context
  of the AIC, the variable IMF model is `decisively preferred for
  $\Delta(\chi^2)>14$; see the text for details.}
\label{fig:dchi2}
\end{figure}

The improvement in the quality of the fits when a variable IMF model
is considered can be quantified by the difference in the minimum of
$\chi^2$.  Figure \ref{fig:dchi2} shows $\Delta(\chi^2)$ between
models that do and do not allow for variation in the IMF as a function
of the difference in the best-fit $M/L$ between the two models.  The
latter quantity is in units of the uncertainty in $M/L$.  Selected
galaxies are labeled in the figure.  There is a clear trend in the
sense that galaxies with mass-to-light ratios that deviate strongly
from what is expected for a MW IMF have large $\Delta(\chi^2)$.  This
of course is not surprising, as it is precisely the $\chi^2$ value
that the fitting routine uses to determine the best-fit $M/L$ ratios.

Quantifying the preference of one model over another is notoriously
complex \citep[see][for a brief review]{Liddle07}.  Among the most
popular metrics are the Akaike and Bayesian information criteria
\citep[the AIC and BIC, respectively;][]{Akaike74, Schwarz78}.  The
former is defined as AIC$\equiv\chi^2+2k$ where $k$ is the number of
free parameters.  The AIC attempts to balance a change in $\chi^2$
against increased model complexity.  In our case the two models under
consideration differ by two parameters, and so an equivalence in AIC
would correspond to $\Delta(\chi^2)=4$.  According to Jeffreys' scale,
$\Delta({\rm AIC})>5$ is judged as `strong' evidence for one model
over another, while $\Delta({\rm AIC})>10$ is judged as `decisive'
\citep{Liddle07}.  We can therefore associate $\Delta(\chi^2)>14$ as a
threshold for decisive preference for the model with IMF variation
over the model without IMF variation.  This threshold is included in
Figure \ref{fig:dchi2}.  Notice that every galaxy with $>3\sigma$
evidence for an IMF different from that of the MW has
$\Delta(\chi^2)>14$, indicating that, in the context of our model, the
variable IMF model is very strongly (decisively) preferred.  We have
not considered the BIC herein because it depends on the number of data
points and assumes that the data points are independent.  This
condition is clearly violated in our case because the velocity
dispersion of each galaxy is larger than the wavelength sampling;
adjacent data points are therefore highly correlated.

\begin{figure*}[!t]
\center
\resizebox{7.5in}{!}{\includegraphics{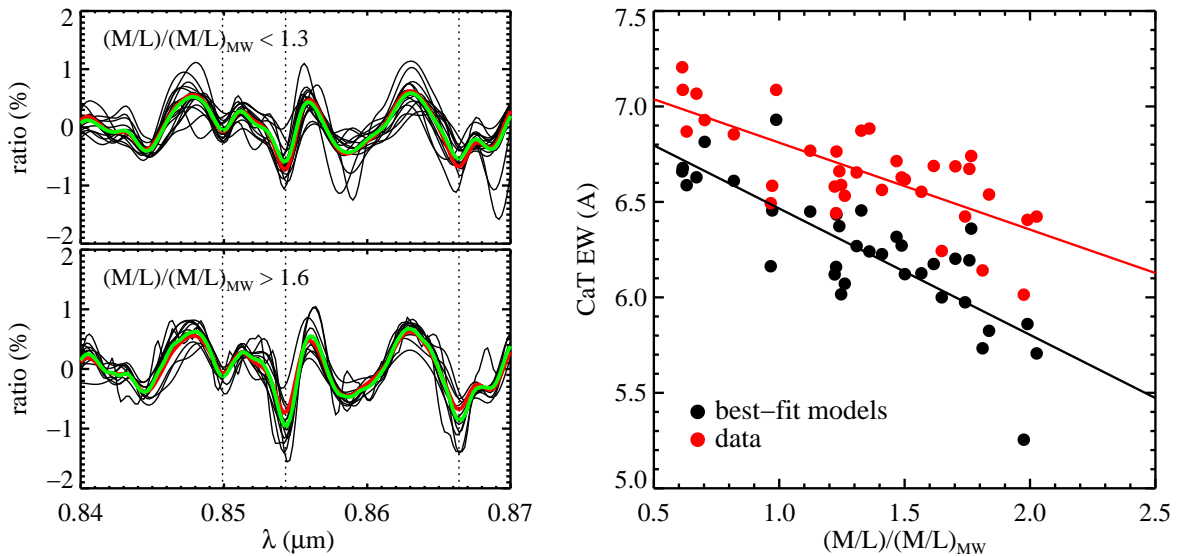}}
\vspace{0.2cm}
\caption{Comparison of models and data around the CaT feature.  {\it
    Left Panel:} Residuals between the best-fit model and data for
  galaxies split according to their best-fit IMF (light IMFs in the
  top panel, heavy IMFs in the bottom panel).  The green lines are the
  mean residuals within each panel, and the red line is the mean
  residual for the entire sample.  Dotted lines mark the location of
  the CaT features.  {\it Right Panel:} Comparison of the CaT EW
  between the best-fit models and the data as a function of the
  best-fit IMF.  Solid lines are a linear fit to the trends.  Notice
  that while the best-fit models are offset low compared to the data,
  the {\it trend} of CaT strength with the inferred IMF is similar
  between the models and the data.  These panels suggest that there is
  an overall offset in the predicted strength of the CaT feature in
  the current version of our models.}
\label{fig:cat}
\end{figure*}

It is apparent that the observed CaT feature is not very well-fit by
the models in Figures \ref{fig:bestspec} and \ref{fig:bestspec2},
especially in the cores of the lines.  This issue is explored further
in Figure \ref{fig:cat}.  In the left panels we show residuals between
the best-fit model and data around the CaT feature for all 35
galaxies, split according to their $M/L$ ratio.  Also included is the
mean residual for the entire sample (red lines) and the mean residuals
within each panel.  Comparison of the green and red lines in these
panels suggests that the CaT feature is poorly fit for all galaxies,
and the quality of the fit does not vary much with $M/L$.  In other
words, there appears to be an overall offset between the models and
the data, independent of the best-fit IMF.  Recall that the [Ca/Fe]
abundance is included in the model fits, and is well-constrained by
the CaI feature at $0.4227\mu m$.  The offset between the models and
data is therefore unlikely to be a calcium abundance effect.

The tension between the observed and modeled CaT is displayed in
another way in the right panel of Figure \ref{fig:cat}.  Here we show
the equivalent width (EW) of the CaT feature \citep[see][for the
definition of this index]{Conroy12a} for both the models and the data
as a function of the best-fit $M/L$ ratio.  The overall trend of
decreasing CaT strength with increasing $M/L$ ratio is expected
because the CaT is a giant-sensitive feature.  The offset between the
models and data is also apparent, and the offset is approximately
constant with $M/L$.  The origin of this offset is unclear.  It may be
due to the fact that the CaT absorption lines form at very low
Rosseland optical depths, and so the model becomes sensitive to the
treatment of the tenuous outer atmospheres of model giants.  It may
also reflect an underlying error in the construction of the base set
of empirical models onto which the synthetic spectra are grafted.
Regardless of the origin, the most important point is that the offset
is almost completely independent of $M/L$, and implies that this
tension between the model and data is unlikely to be driving the
qualitative variation in the IMF that we observe.

\vspace{1cm}

\subsection{The Nuclear Bulge of M31}

The form of the IMF in the nuclear bulge of M31 was the subject of
intense debate in the 1970s and 1980s.  \citet{Spinrad71} observed the
NaI $0.82\mu m$ feature and concluded that the bulge was
dwarf-dominated ($(M/L)_{\rm stars}=44$).  A large number of
subsequent observations have lead to conflicting results
\citep[e.g.,][]{Oconnell76, Whitford77, Frogel78, Cohen78, Faber80,
  Carter86, Delisle92}.  There is broad agreement that the NaI feature
is quite strong, especially in the nuclear region (the inner
$2-4\arcsec$).  The debate centered on whether this enhancement is due
primarily to an increase in the dwarf star contribution or an increase
in the [Na/Fe] abundance.  If interpreted as the former, the implied
mass-to-light ratio would be $M/L_B\approx28$ \citep{Faber80}, which
would be in strong conflict with dynamical constraints
\citep{Saglia10}.

In light of the controversial history of the nuclear bulge of M31 we
decided to include M31 in our analysis.  The LRIS spectrum of M31 was
extracted within a radius of $4\arcsec$, which corresponds to the
central 15 pc of M31.  We fit our SPS model to this spectrum in a
manner identical to the early-type galaxy data.  The resulting
best-fit $\mlk$ is shown as a blue symbol in Figure \ref{fig:imf1}.
We derive a mass-to-light ratio that is in between a MW IMF and a
Salpeter IMF, and is highly inconsistent with dwarf-rich
(bottom-heavy) IMFs.  As will be discussed in Section \ref{s:dyn}, our
derived mass-to-light ratio for M31 agrees well with dynamical
constraints \citep{Saglia10}.

The nuclear spectrum shows an abundance pattern typical of the
early-type galaxy sample.  It is $\alpha-$enhanced in all the $\alpha$
elements except for Ca and Ti, and also enhanced in N and C.  The
derived sodium abundance is however one of the highest of the sample,
at [Na/Fe]$\approx1.0$.  More detailed models and coverage of the NaD
spectral feature will shed further light on the sodium abundance in
M31 and other galaxies.  Removal of the NaI feature results in a
change in $\mlk$ of $-50\%$, while removal of both the NaI and CaT
features results in a change in $\mlk$ of $+50\%$.  These changes are
typical of the systematic uncertainties in our present modeling
technique (see Section \ref{s:sys} for details).

The basic result from this section is that the nuclear region of M31
is entirely consistent with a normal IMF.  We can certainly rule out
extreme mass-to-light ratios, such as those suggested by
\citet{Faber80} based on data obtained within a slightly larger
aperture ($2\arcsec\times4\arcsec$ compared to an $8\arcsec$ circular
diameter herein).  We derive much lower $M/L$ ratios compared to Faber
\& French in part because we allow for large [Na/Fe] variation
(reaching nearly 1 dex for the nuclear spectrum of M31), and in part
because the FeH feature places a strong upper limit on the allowed
$M/L$ values.

\subsection{Metal-Rich Globular Clusters in M31}
\label{s:m31}

We now turn to constraints on $M/L$ for four metal-rich globular
clusters (GCs) in M31.  These clusters were selected to be metal-rich,
$\alpha-$enhanced, and old \citep{Caldwell11}.  They have stellar
masses in the range $5.5<{\rm log}\,M/\Msun<6.1$, based on dynamical
constraints, and long relaxation times \citep{Strader11}.  One of the
most intriguing facts of these clusters is that their $K-$band
mass-to-light ratios are a factor of $\sim2$ {\it lighter} than
expected for a MW IMF \citep{Strader11}.  GCs are simpler than
galaxies in the sense that the stars within them are approximately
coeval, and they are known to contain little or no dark matter
\citep[][and references therein]{Conroy11a}.  Comparison between
dynamical and stellar population-based constraints on $M/L$ is
therefore straightforward.

\begin{figure}[!t]
\center
\resizebox{3.5in}{!}{\includegraphics{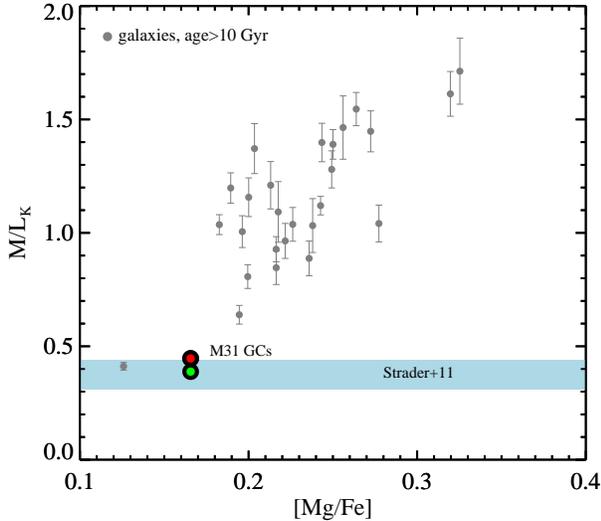}}
\caption{$\mlk$ versus [Mg/Fe] for the early-type galaxies analyzed in
  the present work (grey symbols) and for the stacked spectrum of four
  metal-rich globular clusters in M31 (red symbol; statistical error
  bars are smaller than the symbol size).  We also compute and include
  in the figure the mass-to-light ratio computed without black hole
  and neutron star stellar remnants (green symbol).  The
  dynamically-based $\mlk$ for the four M31 clusters is shown as a
  solid band \citep{Strader11}.  The $\mlk$ for a 13 Gyr, solar
  metallicity MW IMF model is 0.9, implying that these globular
  clusters are {\it lighter} in mass than expected for a MW IMF, and
  consistent with dynamical constraints.}
\label{fig:imfgc}
\end{figure}

We have fit the stacked spectrum of these four clusters with our
model.  The resulting best-fit $K-$band mass-to-light ratio is shown
in Figure \ref{fig:imfgc}, plotted versus the best-fit [Mg/Fe]
abundance.  We also include our sample of early-type galaxies that
have ages $>10$ Gyr.

The standard model $\mlk$ ratio, which includes a contribution from
stellar remnants, is shown as a red symbol. We have also computed a
mass-to-light ratio without inclusion of black hole and neutron star
stellar remnants (green symbol).  A paucity of remnants in GCs may
arise because the shallow potential well of the GCs permits the escape
of massive remnants when the progenitor star explodes.  In any event,
the effect of black holes and neutron stars on the mass-to-light ratio
is small.

Dynamical mass-to-light ratios have been computed for these clusters
by \citet{Strader11}; the full range spanned by these four clusters is
shown as a band in Figure \ref{fig:imfgc}.  For reference, a 13 Gyr,
solar metallicity MW IMF model has $\mlk=0.9$.

Both the stellar population-based and dynamically-based mass-to-light
ratios point toward a bottom-light IMF in these GCs.  This broad
agreement between two completely independent techniques is very
encouraging, and lends support to our model results.

\section{Tests}
\label{s:tests}

In this section we discuss several tests of our principle result that
the IMF is not universal.  We confront our best-fit models with
dynamical constraints in Section \ref{s:dyn} and we discuss several
possible systematics in the model in Section \ref{s:sys}.

\begin{figure}[!t]
\center
\resizebox{3.5in}{!}{\includegraphics{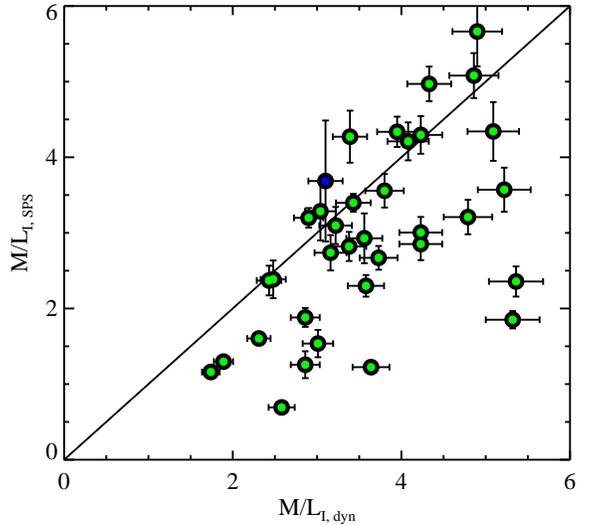}}
\caption{Comparison between SPS-based and dynamically-based $\mli$.
  Notice that the SPS-based mass-to-light ratios never violate the
  dynamical constraints.  M31 is shown separately (blue symbol)
  because the dynamical constraints on $M/L$ come from a different
  source and technique than the rest of the sample.  The $M/L_{\rm
    I,SPS}$ ratio for a MW and a Salpeter IMF is 2.2 and 3.4,
  respectively, assuming an age of 13 Gyr and solar metallicity.}
\label{fig:dyn}
\end{figure}

\subsection{Comparison to Dynamical Masses}
\label{s:dyn}

In this section we compare mass-to-light ratios estimated from our
models to dynamical estimates.  For the 34 galaxies drawn from the
SAURON survey, dynamical mass-to-light ratios are based on Jeans
axisymmetric modeling of the 2D velocity field \citep{Scott09}.  For
these galaxies the dynamical estimates are quoted within $R_e$.  We
have adopted a typical uncertainty on the dynamical $M/L$ values of
6\% following \citet{Cappellari06}.

The comparison between our stellar population-based mass-to-light
ratio, $M/L_{\rm SPS}$, and the dynamically-based $M/L_{\rm dyn}$ is
shown in Figure \ref{fig:dyn}.  Mass-to-light ratios are quoted in the
$I-$band in order to provide a direct comparison with results from the
SAURON survey \citep{Scott09}.  Notice that the SPS and
dynamically-based mass-to-light ratios are obtained within different
physical apertures ($R_e/8$ for the former and $R_e$ for the latter).

\citet{Saglia10} have estimated the dynamical mass-to-light ratio for
M31 in the $R-$band based on long-slit data.  We have assumed
($M/L_R$)/($M/L_I$)=1.3, appropriate for a 13 Gyr single-age
population with solar metallicity, in order to translate their value
into $M/L_I$.  Saglia et al. provide $M/L_R$ as a function of radius;
we adopt a dynamical mass-to-light ratio within $4\arcsec$ for
comparison to our SPS-based value.

The SPS-based $M/L$ values are always less than or equal to the
dynamical $M/L$ values within the $2\sigma$ errors, which is
remarkable given the small size of the SPS-based errors.  Even in
cases where we infer bottom-heavy IMFs, the derived $M/L$ values do
not exceed the dynamical constraints.  There are galaxies for which
the SPS-based $M/L$ ratios are less than the dynamically-based ones.
In these cases there is room for additional, non-stellar mass (i.e.,
dark matter).

\begin{figure*}[!t]
\center
\resizebox{7in}{!}{\includegraphics{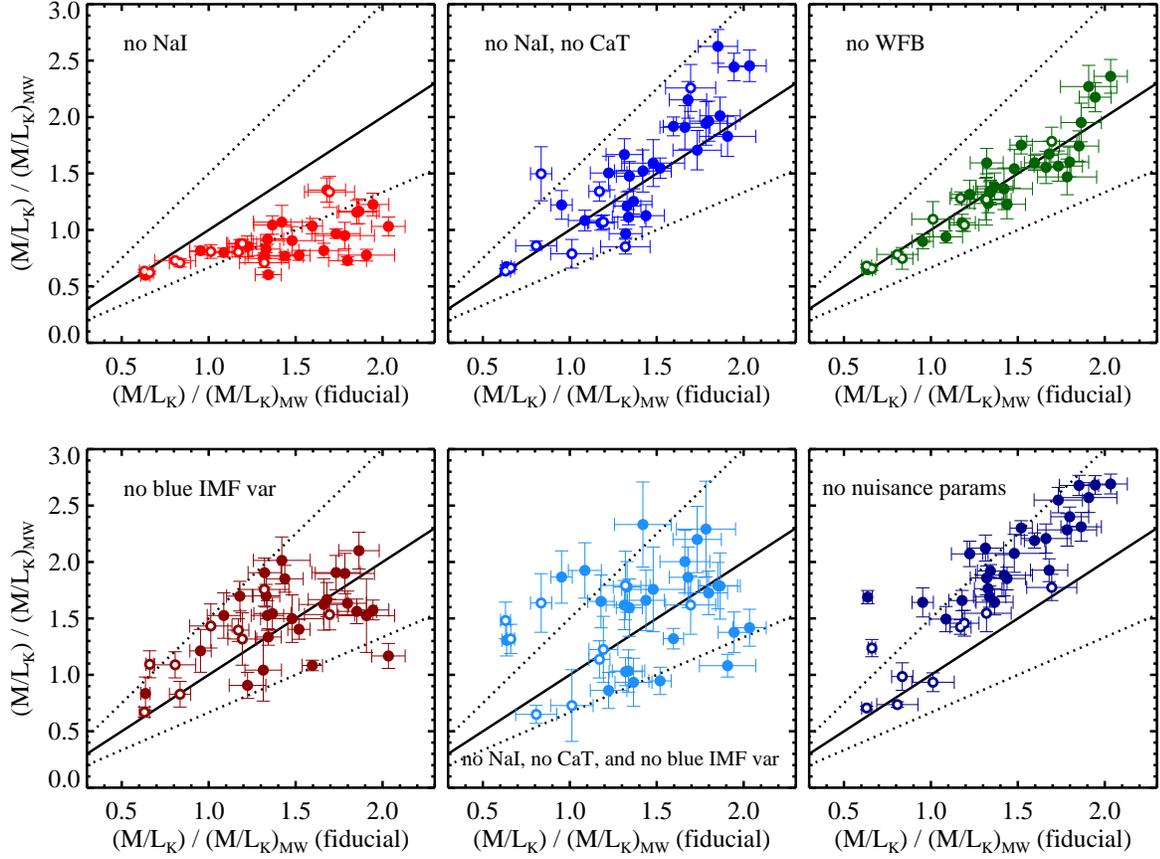}}
\caption{Exploration of systematic effects on the derived
  mass-to-light ratios.  In each panel, the $\mlk$ is plotted in units
  of $(M/L_K)_{\rm MW}$ (as in previous figures).  Dashed lines
  indicate a slope change of $\pm50$\%.  The open symbols in each
  panel indicate galaxies with ages younger than 10 Gyr.  The upper
  panels highlight the effect of removing one or more spectral regions
  from the fit.  In the left panel the NaI feature is removed, in the
  middle panel both the NaI and CaT features are removed, and in the
  right panel the Wing-Ford band spectral region is removed.
  Neglecting certain IMF-sensitive features affects the overall
  normalization of the $\mlk$ values at the $\approx50\%$ level, but
  relative trends are robust.  The lower panels highlight other model
  permutations.  In the left panel we turn off the spectral variation
  due to the IMF in the blue spectral region ($\lambda<0.8\mu m$), in
  the middle panel we turn off blue IMF variation and remove NaI and
  CaT from the fits, and in the right panel the three nuisance
  parameters are not included in the fit.  In the middle panel the
  only constraints on the IMF are coming from the Wing-Ford band
  spectral region.  These tests suggest that the absolute $\mlk$
  ratios carry systematic uncertainties at the $\lesssim50$\% level,
  and that the Wing-Ford band {\it alone} is likely not sufficient to
  place strong constraints on the IMF over the range of IMFs probed.}
\label{fig:sys}
\end{figure*}

\subsection{Systematics}
\label{s:sys}

In this section we explore several variations to our fiducial model.
The variations are of two types: changes to the spectral range
included in the fit, and changes to the properties of the stellar
population model.  We consider the effect of these variations on the
derived $\mlk$ ratios.

We begin by considering variation in the spectral range included in
the fits.  In the top panels of Figure \ref{fig:sys} we consider the
effect on $\mlk$ of removing one or more classic IMF-sensitive
spectral features.  In the left panel we remove the NaI spectral
region ($0.81\mu m-0.83\mu m$).  In the middle panel we remove both
the NaI and CaT features ($0.81\mu m-0.83\mu m$ and $0.845\mu
m-0.87\mu m$).  In this case the Wing-Ford band is the only classic
IMF-sensitive feature included in the fit.  In the right panel we
remove from the fit all information beyond $\lambda>0.96\mu m$, which
includes the Wing-Ford band.

Qualitatively we recover the same trends no matter which set of
IMF-sensitive features are used.  In particular, in all cases we find
evidence for IMF variation.  This provides strong confirmation that
our basic result is not influenced by a single spectroscopic feature.
There are however noticeable offsets between the various permutations,
especially in the upper left panel of Figure \ref{fig:sys}.  This
means that in detail the various IMF-sensitive features are favoring
different values of $\mlk$, which in turn suggests that there are
residual systematics in the modeling.  One possible explanation lies
in the fact that these features are each most sensitive to a different
range of stellar masses (see CvD12 for details).  The data may be
demanding an IMF that is more complicated than the broken power-law
that we adopt.  These issues will be explored in future work.

We have also considered a model where the effect of the IMF on the
spectrum is limited to the red ($\lambda>0.8\mu m$).  In this model
the blue spectrum is completely insensitive to IMF variation.  Our
goal here is to assess the extent to which the IMF spectral signatures
in the blue are driving the IMF constraints.  The result is shown in
the bottom left panel of Figure \ref{fig:sys}.  The $\mlk$ ratios in
this case are again similar to the fiducial model, with a scatter
between the two of 30\%.  The sizable scatter suggests that there is
in fact some information on the IMF in the blue spectral range, which
in retrospect should not be surprising (see e.g., Figure 10 in CvD12).
However, to the extent that the blue spectral region is sensitive to
the IMF, it is almost certainly sensitive to a different stellar mass
range, since the coolest M dwarfs have such red spectral energy
distributions.

In the lower middle panel of Figure \ref{fig:sys} we consider an
extreme model in which the IMF variation in the blue spectral region
is suppressed {\it and} the NaI and CaT spectral features are masked
from the fit.  In this case IMF constraints are due only to the
Wing-Ford band.  The scatter between this model and the fiducial is
large.  Evidently, over the range of IMFs probed in our fiducial
model, the Wing-Ford spectral region is not a particularly powerful
probe of the IMF when considered in isolation.  This is also evident
in Figure 10 of Paper I, which shows that the relation between the
Wing-Ford band and $\sigma$ has considerable scatter.  It can however
be a useful probe of the IMF in conjunction with other features, as
illustrated in the upper panels of this figure.  As discussed
extensively in CvD12, the Wing-Ford band is most sensitive to the very
lowest mass stars, and so the relatively large scatter between the IMF
inferred from the Wing-Ford band alone and the fiducial model may be
pointing toward the need to consider additional flexibility in the
model IMF at low masses.

We also consider variation in the underlying model by removing the
three nuisance parameters in the fiducial model, including log(M7III),
log$(f_{\rm hot})$, and $T_{\rm hot}$.  This is shown in the lower
right panel of Figure \ref{fig:sys}.  Here again the qualitative trend
of IMF variation is robust.  In detail however there is a fair degree
of scatter between this model and the fiducial, and the typical $\mlk$
is $\sim30\%$ higher when the nuisance parameters are removed.  The
most important nuisance parameter driving these differences is
log(M7III).  Galaxies for which this parameter is close to zero have
very similar $\mlk$ regardless of whether or not the nuisance
parameters are included.  In contrast, those galaxies which have the
largest values of log(M7III), of order $-1.0$, show the largest
differences in $\mlk$, of approximately 50\%, when the nuisance
parameters are or are not included.  More detailed modeling of the
cool giant contribution to the observed spectra will be required
before more accurate IMF constraints can be obtained.

The results of this section demonstrate that our primary result, that
the IMF varies from galaxy to galaxy, is robust to a variety of model
permutations.  There are however systematic uncertainties in the
mass-to-light ratios at the $\approx50\%$ level attributable to the
choice of the parameter set and wavelength coverage.  To first order,
this systematic uncertainty applies equally to all mass-to-light
ratios, so that {\it relative} trends are quite robust. Further work
is needed to address the remaining sources of systemtatic uncertainty
in the model.

\section{Discussion}
\label{s:disc}

\subsection{Is IMF Variation Really Necessary?}
\label{s:imfnec}

One of the key goals of this paper has been to demonstrate that IMF
variation is not only plausible but {\it necessary} given the data and
a very flexible SPS model.  We have employed a model that allows for
variation in 11 elements and an additional 6 parameters (not including
the IMF) meant to capture a wide range in possible underlying stellar
populations.  Nonetheless, our model is obviously not exhaustive, and
so one may wonder whether additional components not included in our
model may mimic the effect of IMF variation.  Ultimately this question
must be addressed with further quantitative modeling; in this section
we limit ourselves to more general statements.

The most significant argument favoring IMF variation is that the three
classic IMF-sensitive spectral features --- NaI, CaT, and FeH --- are
all telling a broadly consistent story (as evidenced by the upper
panels of Figure \ref{fig:sys}).  Any model that is proposed to
explain away the purported IMF variation must simultaneously decrease
the strength of CaT, increase the strength of NaI, and also explain
the strength of the Wing-Ford band.  It is not easy to imagine a
single population that can satisfy these constraints.

As an example, cool stars have strong TiO bands throughout their
spectra. In particular, the $2-3$ band of the $\delta$ system of TiO
coincides with the FeH feature at $0.99\mu m$ \citep[see the
supplementary material in][]{vanDokkum10}, while the $1-0$ band of the
$\delta$ system coincides roughly with the NaI feature at $0.82\mu m$.
An increase in the contribution of late M giant light may therefore
increase the strength of the Wing-Ford band and the NaI feature.  We
have included a cool M giant in our model as a nuisance parameter, but
one could argue that other, perhaps more peculiar cool stars exist in
early-type galaxies.  This scenario is however strongly constrained by
the central wavelength of the NaI feature, which becomes
systematically bluer with increasing velocity dispersion (see Paper
I).  The reason for this is because, while the NaI and TiO features
roughly coincide, the former has a bluer central wavelength than the
latter. If TiO was the cause of the strong NaI and Wing-Ford features,
then the centroid of the NaI feature would become {\it redder}, not
bluer with increasing velocity dispersion.  Furthermore, the strength
of TiO bands is constrained from other, cleaner, spectral regions
including the $0-0\,\delta$ TiO bandhead at $0.88-0.89\mu m$.  In
fact, when both the NaI and CaT features are masked, a strong
constraint on the IMF is still obtained because the $0.88-0.89\mu m$
TiO bandhead constrains the possible contribution from TiO to the
Wing-Ford band.

One may also wonder if, rather than an increase in low-mass stars, one
could explain the observations with a {\it decrease} in the number of
giant branch stars.  These two options would be at least approximately
equivalent were it not for the turn-off stars, which provide a
benchmark against which one can constrain {\it both} the relative
number of late-type giants and dwarfs.  Here again the Wing-Ford band
provides a useful diagnostic.  As demonstrated in CvD12, to some
extent (and ignoring constraints from other spectral regions), the
strength of the NaI and CaT features can be equally well fit with an
increase in dwarfs or a decrease in giants.  But this is not true for
the Wing-Ford band, which becomes strong only in the latest M dwarfs
due to FeH absorption.  For a MW IMF, these M dwarfs are too faint to
contribute any signal to the Wing-Ford band, even when luminous giants
are removed from the model.  In fact, decreasing the number of giants
in the model actually decreases the strength of the Wing-Ford band for
a MW IMF.  This occurs because the Wing-Ford band is also influenced
by TiO absorption, which decreases as the number of giants decreases.

\begin{figure*}[!t]
\resizebox{7.5in}{!}{\includegraphics{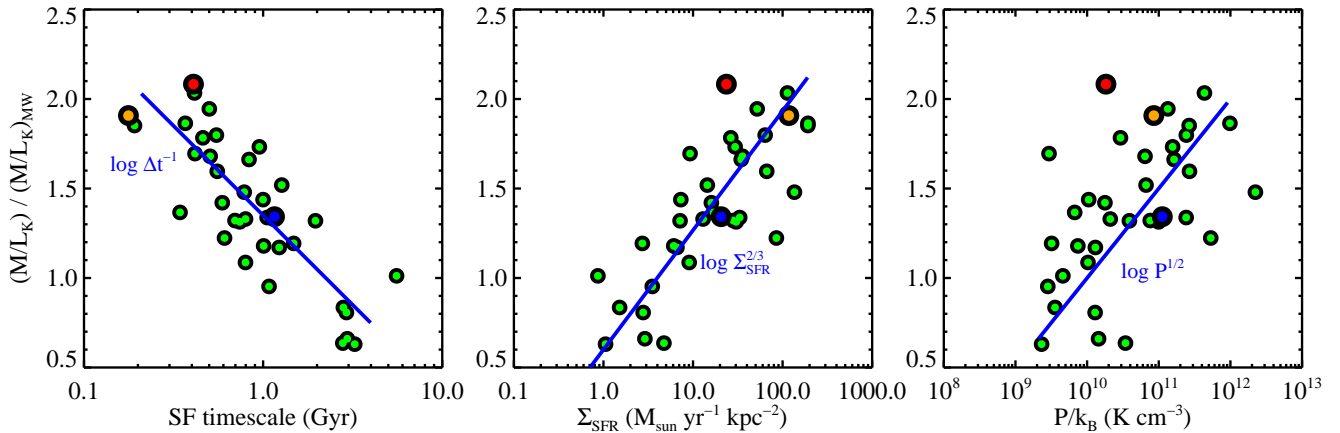}}
\vspace{0.1cm}
\caption{IMF versus inferred SF timescale (left panel), inferred SFR
  surface density, $\Sigma_{\rm SFR}$ (middle panel), and inferred
  pressure (right panel).  All quantities are measured within $R_e/8$
  except for M31, where quantities are measured within the central
  $4\arcsec=15$ pc.  The quantity plotted on the $y-$axis is sensitive
  only to the IMF, as it is the true mass-to-light ratio divided by
  the mass-to-light ratio assuming a MW IMF.  The SF timescale is
  inferred from the measured [Mg/Fe] abundance.  The SFR surface
  density is estimated from the stellar mass, SF timescale, and
  present galaxy size, and the pressure is estimated from the mass and
  present galaxy size.  Symbol colors are as in Figure \ref{fig:imf1}.
  Solid lines are not fits to the points, they are simply intended to
  guide the eye.}
\label{fig:dt}
\end{figure*}

Another concern is related to the reliability of the models at high
metallicity.  Recall that we are not explicitly allowing the
isochrones to vary with metallicity.  Instead we have adopted an
additional free parameter, $\Delta(\teff)$, which allows us to mimic
shifts in the isochrones by varying $\teff$ for all the stars in the
model.  It is important to realize the total variation in metallicity
across our sample is not large: [Z/H] varies from $\approx0.0$ to
$\approx0.25$ with a median of 0.15 (see Figure \ref{fig:imfz}).
There are many galaxies in our sample with [Fe/H]$\approx-0.1$ and
[Mg/Fe]$\approx0.2$ and thus [Z/H]$\approx0.1$.  Using the Dartmouth
Stellar Evolution
Database\footnote{http://stellar.dartmouth.edu/$\sim$models/}, we have
created isochrones with [Fe/H]$=-0.1$ and [Mg/Fe]$=0.2$ and find that
these isochrones are almost identical to the solar metallicity,
solar-scaled isochrones that we adopt for our base model.  The most
metal-rich galaxies in our sample have isochrones that may be $50-100$
K cooler than those used in our base model.  Figure \ref{fig:imfz}
reveals a weak correlation of IMF with [Z/H], and, importantly, at
[Z/H]$>0.1$ the inferred IMFs range from MW-like to steeper than
Salpeter.  This suggests that any model systematics that correlate
with metallicity are not driving the inferred IMF variation with
$\sigma$ and [Mg/Fe] shown in Figure \ref{fig:imf1}.  Ultimately, more
work is needed in the construction and calibration of models at high
metallicity.

It will be difficult to prove unequivocally that the IMF does indeed
vary from galaxy to galaxy based on integrated light measurements.
However, it presently remains the best explanation for the observed
spectral features in early-type galaxies given the available models.

\subsection{Origin of the Observed Trends}
\label{s:origin}

If we now take the inferred IMF variation at face value, we can ask
what physical mechanism(s) may give rise to the observed correlations.
The observed correlation between the IMF and $\alpha-$enhancement
appears to be stronger than the correlation with $\sigma$; we will
therefore interpret the former as the more fundamental relation.  Of
course, nearly all properties of early-type galaxies are strongly
correlated with one another \citep[e.g.,][]{Faber76, Djorgovski87,
  Worthey92, Trager00, Thomas05, Graves09a}, so further work will be
required to positively identify the fundamental underlying variable(s)
governing the variation in the IMF.

The level of $\alpha-$enhancement in a stellar population is normally
interpreted in terms of a star formation (SF) timescale --- higher
$\alpha-$enhancements correspond to shorter SF timescales.  We adopt a
relation between [$\alpha$/Fe] and SF timescale based on a simple
chemical evolution model presented in \citet{Thomas05},
[$\alpha$/Fe]$\approx\frac{1}{5}-\frac{1}{6}{\rm log} \Delta t$, and
in this section we assume that [Mg/Fe]$=$[$\alpha$/Fe].  We caution
that the precise relation between timescale and $\alpha-$enhancement
depends on the stellar population model and the details of galactic
chemical evolution \citep[e.g.,][]{Arrigoni10}.  The resulting relation
between the IMF and SF timescale is shown in the left panel of Figure
\ref{fig:dt}.  Notice that the galaxies with the most bottom-heavy
IMFs have inferred SF timescales of only $200-300$ Myr.

We can go one step further and estimate an average star formation rate
(SFR) surface density, $\Sigma_{\rm SFR}$, within $R_e/8$ based on the
stellar mass within this radius\footnote{The stellar mass within
  $R_e/8$ is estimated by combining the best-fit $\mlk$ ratios, the
  total $K-$band luminosities from \citet{Cappellari11}, and the
  Sersic indices from \citet{FalconBarroso11}.  The latter quantity is
  used to estimate the fraction of the total light contained within
  $R_e/8$.  For M31 the $I-$band luminosity within our extraction
  aperture was derived from Sick et al., in prep.}, $M_\ast$, and the
SF timescale: $\Sigma_{\rm SF}=M_\ast/4\pi\Delta t (R_e/8)^2$.  The
result is shown in the middle panel of Figure \ref{fig:dt}.  Galaxies
with the most bottom-heavy IMF have inferred SFR surface densities in
excess of $100\,\Msun\, {\rm yr}^{-1}\, {\rm kpc}^{-2}$.  In the local
universe such high SFR surface densities are found only in the most
extreme circumnuclear starbursts \citep{Kennicutt98}.  We emphasize
that these quantities are based on the {\it present} stellar density
within $R_e/8$.  Observations of massive early-type galaxies suggest
that their central densities actually decrease with time
\citep{Bezanson09}, which appears to be a consequence of both major
and minor mergers \citep{Oser12}.  These inferred SFR surface
densities may therefore be lower limits to the true SFR densities.

Another quantity of interest is the pressure of the system, which can
be estimated via $P\propto M_\ast^2/R^4$.  We have computed the
effective pressure within $R_e/8$ for our sample and plotted this
against our best-fit IMFs in the right panel of Figure \ref{fig:dt}.
A correlation is apparent, but it is weaker than the other relations
shown in this figure.  This may be due to the fact that the effective
pressure estimated at $z=0$ is only weakly correlated with the pressure
at the epoch of formation.

In each of these panels we have included simple power-law relations.
These were not fits to the data; the power-law indices were chosen by
eye to represent the mean trend in the sample.

The data therefore support a scenario wherein the IMF is correlated
with the intensity of star formation and/or the effective pressure of
the system, in the sense that higher SFR densities and higher
pressures correspond to more bottom-heavy IMFs.

A number of recent papers have pointed to fragmentation in
supersonically turbulent molecular clouds as the key physical process
governing the shape of the stellar IMF \citep{Padoan97, Padoan02,
  Hennebelle08, Hopkins12a, Hopkins12b}.  In this framework the Mach
number within the molecular cloud is a key variable affecting the
shape of the IMF.  Qualitatively, and for all other variables held
fixed, a higher Mach number results in a more bottom-heavy IMF.

As noted above, local circumnuclear starbursts may be analogs to the
most massive early-type galaxies in our sample at their formation
epoch.  The local starbursts have very high SFR densities and have
molecular gas that is on average hotter and more supersonically
turbulent than the molecular gas in the Milky Way \citep{Downes98,
  Bryant99}.  The higher Mach numbers in starbursts may be the result
of the higher rate of supernovae (SNe) in these systems.
\citet{Hopkins12b} has combined these ideas with an analytic theory
for the IMF to demonstrate that such galaxies should have bottom-heavy
IMFs.  The Mach number is critical to this conclusion: the thermal
Jeans mass ($M_J\propto c_s^3\,\rho^{-1/2}$, where $c_s$ and $\rho$
are the sound speed and density, respectively) in starbursts is
frequently larger than in the MW disk, and thus a simple Jeans
argument would suggest that the IMF in starbursts should be {\it
  bottom-light} \citep[e.g.,][]{Larson98, Baugh05, Narayanan12}.  The
larger Mach numbers in starbursts overcomes the effect of the
increasing thermal Jeans mass to result in a lower characteristic mass
in the IMF.

The basic idea is that galaxies with higher sustained $\Sigma_{\rm
  SFR}$ have a rate per unit volume of SNe, which can drive more
highly supersonic turbulence, resulting in a higher typical Mach
number in molecular clouds that in turn lowers the characteristic
mass of the IMF.  Notice that this picture does not imply that all
$\alpha-$enhanced systems have bottom-heavy IMFs.  The $\Sigma_{\rm
  SFR}$ must be sustained long enough for the SNe to drive turbulence.
In particular, globular clusters, which are $\alpha-$enhanced systems,
would not be expected to have bottom-heavy IMFs because star formation
in such systems is approximately instantaneous.

We conclude this section with the caution that isothermality, a key
assumption in the models mentioned above, likely does not hold in the
regimes of interest.  \citet{Krumholz11} has argued instead that
radiative feedback is the key physical process setting the
characteristic mass of the IMF, and that the Jeans mass (whether
thermal or turbulent) plays no direct role.  In Krumholz's model the
ISM pressure is the key variable, and there may be some evidence for
IMF variation with pressure in Figure \ref{fig:dt}.

\section{Summary}
\label{s:sum}

In this paper we have confronted high-quality absorption line spectra
of 38 early-type galaxies and the nuclear bulge of M31 with a new
stellar population synthesis model that incorporates flexible
abundance patterns and IMFs.  These data and models extend beyond
$1\mu m$, where IMF-sensitive absorption features allow for a strong
constraint on the IMF and stellar mass-to-light ratio within
individual galaxies.  The data sample the inner regions of the
galaxies (to $R_e/8$) and so conclusions regarding the stellar
populations of these galaxies apply strictly to these inner regions.
We now summarize our main results.

\begin{itemize}

\item Evidence is found for an IMF that varies systematically with
  galaxy velocity dispersion and $\alpha-$enhancement.  Steeper (more
  bottom-heavy) IMFs are found in more massive systems.  The best-fit
  mass-to-light ratios do not violate dynamical constraints.

\item At the highest velocity dispersions and $\alpha-$enhancements
  the IMF becomes steeper than even the canonical Salpeter IMF, with
  inferred $K-$band mass-to-light ratios a factor of $\approx2$ higher
  than would be expected for a universal, Milky Way IMF.

\item Systematic uncertainties in the models translate into
  $\lesssim50\%$ uncertainties in the derived mass-to-light ratios,
  while the median statistical uncertainty is $\approx7\%$.  In
  particular, there appears to be some tension between the models and
  data regarding the CaT feature, which is interesting because CaT is
  strong in giants while the others are strong in dwarfs.  We have
  demonstrated that the tension around the CaT feature is visible for
  all of the galaxies in our sample and suggests that the relatively
  poor modeling of the CaT feature does not strongly influence the
  derived $M/L$ ratios.

\item These results are consistent with a scenario wherein the IMF
  becomes increasingly bottom-heavy as the SF timescale becomes
  increasingly short, the SFR surface density becomes increasingly
  high, and/or the ISM pressure becomes increasingly high.  These
  trends are broadly consistent with several recent conjectures for
  the origin of the IMF, but more detailed models are needed before
  conclusive statements can be made.

\end{itemize}


\acknowledgments 

We thank Nelson Caldwell for providing his Hectospec data for the M31
globular clusters, Ricardo Schiavon for assistance with interpreting
his models, Michele Cappellari for useful discussions regarding
uncertainties in SAURON $M/L$ values, and Jonathan Sick and St\'ephane
Courteau for providing the nuclear luminosity of M31 from their
unpublished data.  CC thanks Phil Hopkins, Andrey Kravtsov and Mark
Krumholz for informative discussions.  We acknowledge use of the
Odyssey cluster supported by the FAS Science Division Research
Computing Group at Harvard University.  Finally, we thank the referee
for thoughtful comments that have improved the quality of the
manuscript.

The data presented herein were obtained at the W. M. Keck Observatory,
which is operated as a scientific partnership among the California
Institute of Technology, the University of California and the National
Aeronautics and Space Administration.  The Observatory was made
possible by the generous financial support of the W. M. Keck
Foundation. The authors wish to recognize and acknowledge the very
significant cultural role and reverence that the summit of Mauna Kea
has always had within the indigenous Hawaiian community.  We are most
fortunate to have the opportunity to conduct observations from this
mountain.


\begin{thebibliography}{87}
\expandafter\ifx\csname natexlab\endcsname\relax\def\natexlab#1{#1}\fi

\bibitem[{{Adams} \& {Fatuzzo}(1996)}]{Adams96}
{Adams}, F.~C. \& {Fatuzzo}, M. 1996, \apj, 464, 256

\bibitem[{{Akaike}(1974)}]{Akaike74}
{Akaike}, H. 1974, IEEE Transactions on Automatic Control, 19, 716

\bibitem[{{Arrigoni} {et~al.}(2010){Arrigoni}, {Trager}, {Somerville}, \&
  {Gibson}}]{Arrigoni10}
{Arrigoni}, M., {Trager}, S.~C., {Somerville}, R.~S., \& {Gibson}, B.~K. 2010,
  \mnras, 402, 173

\bibitem[{{Auger} {et~al.}(2010){Auger}, {Treu}, {Gavazzi}, {Bolton},
  {Koopmans}, \& {Marshall}}]{Auger10}
{Auger}, M.~W., {Treu}, T., {Gavazzi}, R., {Bolton}, A.~S., {Koopmans},
  L.~V.~E., \& {Marshall}, P.~J. 2010, \apjl, 721, L163

\bibitem[{{Bacon} {et~al.}(2001)}]{Bacon01}
{Bacon}, R. {et~al.} 2001, \mnras, 326, 23

\bibitem[{{Baraffe} {et~al.}(1998){Baraffe}, {Chabrier}, {Allard}, \&
  {Hauschildt}}]{Baraffe98}
{Baraffe}, I., {Chabrier}, G., {Allard}, F., \& {Hauschildt}, P.~H. 1998, \aap,
  337, 403

\bibitem[{{Bastian} {et~al.}(2010){Bastian}, {Covey}, \& {Meyer}}]{Bastian10}
{Bastian}, N., {Covey}, K.~R., \& {Meyer}, M.~R. 2010, \araa, 48, 339

\bibitem[{{Baugh} {et~al.}(2005){Baugh}, {Lacey}, {Frenk}, {Granato}, {Silva},
  {Bressan}, {Benson}, \& {Cole}}]{Baugh05}
{Baugh}, C.~M., {Lacey}, C.~G., {Frenk}, C.~S., {Granato}, G.~L., {Silva}, L.,
  {Bressan}, A., {Benson}, A.~J., \& {Cole}, S. 2005, \mnras, 356, 1191

\bibitem[{{Bezanson} {et~al.}(2009){Bezanson}, {van Dokkum}, {Tal},
  {Marchesini}, {Kriek}, {Franx}, \& {Coppi}}]{Bezanson09}
{Bezanson}, R., {van Dokkum}, P.~G., {Tal}, T., {Marchesini}, D., {Kriek}, M.,
  {Franx}, M., \& {Coppi}, P. 2009, \apj, 697, 1290

\bibitem[{{Bryant} \& {Scoville}(1999)}]{Bryant99}
{Bryant}, P.~M. \& {Scoville}, N.~Z. 1999, \aj, 117, 2632

\bibitem[{{Caldwell} {et~al.}(2011){Caldwell}, {Schiavon}, {Morrison}, {Rose},
  \& {Harding}}]{Caldwell11}
{Caldwell}, N., {Schiavon}, R., {Morrison}, H., {Rose}, J.~A., \& {Harding}, P.
  2011, \aj, 141, 61

\bibitem[{{Cappellari} {et~al.}(2006)}]{Cappellari06}
{Cappellari}, M. {et~al.} 2006, \mnras, 366, 1126

\bibitem[{{Cappellari} {et~al.}(2011)}]{Cappellari11}
---. 2011, \mnras, 413, 813

\bibitem[{{Cappellari} {et~al.}(2012)}]{Cappellari12}
---. 2012, \nat, 484, 485

\bibitem[{{Carter} {et~al.}(1986){Carter}, {Visvanathan}, \&
  {Pickles}}]{Carter86}
{Carter}, D., {Visvanathan}, N., \& {Pickles}, A.~J. 1986, \apj, 311, 637

\bibitem[{{Cenarro} {et~al.}(2003){Cenarro}, {Gorgas}, {Vazdekis}, {Cardiel},
  \& {Peletier}}]{Cenarro03}
{Cenarro}, A.~J., {Gorgas}, J., {Vazdekis}, A., {Cardiel}, N., \& {Peletier},
  R.~F. 2003, \mnras, 339, L12

\bibitem[{{Chabrier}(2003)}]{Chabrier03}
{Chabrier}, G. 2003, \pasp, 115, 763

\bibitem[{{Chabrier} \& {Baraffe}(1997)}]{Chabrier97}
{Chabrier}, G. \& {Baraffe}, I. 1997, \aap, 327, 1039

\bibitem[{{Coelho} {et~al.}(2007){Coelho}, {Bruzual}, {Charlot}, {Weiss},
  {Barbuy}, \& {Ferguson}}]{Coelho07}
{Coelho}, P., {Bruzual}, G., {Charlot}, S., {Weiss}, A., {Barbuy}, B., \&
  {Ferguson}, J.~W. 2007, \mnras, 382, 498

\bibitem[{{Cohen}(1978)}]{Cohen78}
{Cohen}, J.~G. 1978, \apj, 221, 788

\bibitem[{{Conroy} {et~al.}(2009){Conroy}, {Gunn}, \& {White}}]{Conroy09a}
{Conroy}, C., {Gunn}, J.~E., \& {White}, M. 2009, \apj, 699, 486

\bibitem[{{Conroy} {et~al.}(2011){Conroy}, {Loeb}, \& {Spergel}}]{Conroy11a}
{Conroy}, C., {Loeb}, A., \& {Spergel}, D.~N. 2011, \apj, 741, 72

\bibitem[{{Conroy} \& {van Dokkum}(2012)}]{Conroy12a}
{Conroy}, C. \& {van Dokkum}, P. 2012, \apj, 747, 69

\bibitem[{{Couture} \& {Hardy}(1993)}]{Couture93}
{Couture}, J. \& {Hardy}, E. 1993, \apj, 406, 142

\bibitem[{{Cushing} {et~al.}(2005){Cushing}, {Rayner}, \& {Vacca}}]{Cushing05}
{Cushing}, M.~C., {Rayner}, J.~T., \& {Vacca}, W.~D. 2005, \apj, 623, 1115

\bibitem[{{de Zeeuw} {et~al.}(2002)}]{deZeeuw02}
{de Zeeuw}, P.~T. {et~al.} 2002, \mnras, 329, 513

\bibitem[{{Delisle} \& {Hardy}(1992)}]{Delisle92}
{Delisle}, S. \& {Hardy}, E. 1992, \aj, 103, 711

\bibitem[{{Djorgovski} \& {Davis}(1987)}]{Djorgovski87}
{Djorgovski}, S. \& {Davis}, M. 1987, \apj, 313, 59

\bibitem[{{Dotter} {et~al.}(2007){Dotter}, {Chaboyer}, {Ferguson}, {Lee},
  {Worthey}, {Jevremovi{\'c}}, \& {Baron}}]{Dotter07}
{Dotter}, A., {Chaboyer}, B., {Ferguson}, J.~W., {Lee}, H.-c., {Worthey}, G.,
  {Jevremovi{\'c}}, D., \& {Baron}, E. 2007, \apj, 666, 403

\bibitem[{{Dotter} {et~al.}(2008){Dotter}, {Chaboyer}, {Jevremovi{\'c}},
  {Kostov}, {Baron}, \& {Ferguson}}]{Dotter08b}
{Dotter}, A., {Chaboyer}, B., {Jevremovi{\'c}}, D., {Kostov}, V., {Baron}, E.,
  \& {Ferguson}, J.~W. 2008, \apjs, 178, 89

\bibitem[{{Downes} \& {Solomon}(1998)}]{Downes98}
{Downes}, D. \& {Solomon}, P.~M. 1998, \apj, 507, 615

\bibitem[{{Dunkley} {et~al.}(2005){Dunkley}, {Bucher}, {Ferreira}, {Moodley},
  \& {Skordis}}]{Dunkley05}
{Dunkley}, J., {Bucher}, M., {Ferreira}, P.~G., {Moodley}, K., \& {Skordis}, C.
  2005, \mnras, 356, 925

\bibitem[{{Dutton} {et~al.}(2011){Dutton}, {Conroy}, {van den Bosch}, {Simard},
  {Mendel}, {Courteau}, {Dekel}, {More}, \& {Prada}}]{Dutton11b}
{Dutton}, A.~A., {Conroy}, C., {van den Bosch}, F.~C., {Simard}, L., {Mendel},
  J.~T., {Courteau}, S., {Dekel}, A., {More}, S., \& {Prada}, F. 2011, \mnras,
  416, 322

\bibitem[{{Dutton} {et~al.}(2012{\natexlab{a}}){Dutton}, {Maccio'}, {Mendel},
  \& {Simard}}]{Dutton12b}
{Dutton}, A.~A., {Maccio'}, A.~V., {Mendel}, J.~T., \& {Simard}, L.
  2012{\natexlab{a}}, ArXiv:1204.2825

\bibitem[{{Dutton} {et~al.}(2012{\natexlab{b}}){Dutton}, {Mendel}, \&
  {Simard}}]{Dutton12a}
{Dutton}, A.~A., {Mendel}, J.~T., \& {Simard}, L. 2012{\natexlab{b}}, \mnras,
  422, L33

\bibitem[{{Faber} \& {French}(1980)}]{Faber80}
{Faber}, S.~M. \& {French}, H.~B. 1980, \apj, 235, 405

\bibitem[{{Faber} \& {Jackson}(1976)}]{Faber76}
{Faber}, S.~M. \& {Jackson}, R.~E. 1976, \apj, 204, 668

\bibitem[{{Falc{\'o}n-Barroso} {et~al.}(2011)}]{FalconBarroso11}
{Falc{\'o}n-Barroso}, J. {et~al.} 2011, \mnras, 417, 1787

\bibitem[{{Frogel} {et~al.}(1978){Frogel}, {Persson}, {Matthews}, \&
  {Aaronson}}]{Frogel78}
{Frogel}, J.~A., {Persson}, S.~E., {Matthews}, K., \& {Aaronson}, M. 1978,
  \apj, 220, 75

\bibitem[{{Graves} {et~al.}(2009){Graves}, {Faber}, \& {Schiavon}}]{Graves09a}
{Graves}, G.~J., {Faber}, S.~M., \& {Schiavon}, R.~P. 2009, \apj, 693, 486

\bibitem[{{Graves} \& {Schiavon}(2008)}]{Graves08}
{Graves}, G.~J. \& {Schiavon}, R.~P. 2008, \apjs, 177, 446

\bibitem[{{Grillo} \& {Gobat}(2010)}]{Grillo10}
{Grillo}, C. \& {Gobat}, R. 2010, \mnras, 402, L67

\bibitem[{{Grillo} {et~al.}(2008){Grillo}, {Gobat}, {Rosati}, \&
  {Lombardi}}]{Grillo08}
{Grillo}, C., {Gobat}, R., {Rosati}, P., \& {Lombardi}, M. 2008, \aap, 477, L25

\bibitem[{{Hardy} \& {Couture}(1988)}]{Hardy88}
{Hardy}, E. \& {Couture}, J. 1988, \apjl, 325, L29

\bibitem[{{Hennebelle} \& {Chabrier}(2008)}]{Hennebelle08}
{Hennebelle}, P. \& {Chabrier}, G. 2008, \apj, 684, 395

\bibitem[{{Hopkins}(2012{\natexlab{a}})}]{Hopkins12a}
{Hopkins}, P.~F. 2012{\natexlab{a}}, \mnras, 423, 2037

\bibitem[{{Hopkins}(2012{\natexlab{b}})}]{Hopkins12b}
---. 2012{\natexlab{b}}, ArXiv:1204.2835

\bibitem[{{Kennicutt}(1998)}]{Kennicutt98}
{Kennicutt}, Jr., R.~C. 1998, \apj, 498, 541

\bibitem[{{Kroupa}(2001)}]{Kroupa01}
{Kroupa}, P. 2001, \mnras, 322, 231

\bibitem[{{Kroupa} {et~al.}(2011){Kroupa}, {Weidner}, {Pflamm-Altenburg},
  {Thies}, {Dabringhausen}, {Marks}, \& {Maschberger}}]{Kroupa12}
{Kroupa}, P., {Weidner}, C., {Pflamm-Altenburg}, J., {Thies}, I.,
  {Dabringhausen}, J., {Marks}, M., \& {Maschberger}, T. 2011, ArXiv:1112.3340

\bibitem[{{Krumholz}(2011)}]{Krumholz11}
{Krumholz}, M.~R. 2011, \apj, 743, 110

\bibitem[{{Kuntschner} {et~al.}(2010)}]{Kuntschner10}
{Kuntschner}, H. {et~al.} 2010, \mnras, 408, 97

\bibitem[{{Larson}(1998)}]{Larson98}
{Larson}, R.~B. 1998, \mnras, 301, 569

\bibitem[{{Larson}(2005)}]{Larson05}
---. 2005, \mnras, 359, 211

\bibitem[{{Lee} {et~al.}(2009){Lee}, {Worthey}, {Dotter}, {Chaboyer},
  {Jevremovi{\'c}}, {Baron}, {Briley}, {Ferguson}, {Coelho}, \&
  {Trager}}]{LeeHC09}
{Lee}, H.-c., {Worthey}, G., {Dotter}, A., {Chaboyer}, B., {Jevremovi{\'c}},
  D., {Baron}, E., {Briley}, M.~M., {Ferguson}, J.~W., {Coelho}, P., \&
  {Trager}, S.~C. 2009, \apj, 694, 902

\bibitem[{{Liddle}(2007)}]{Liddle07}
{Liddle}, A.~R. 2007, \mnras, 377, L74

\bibitem[{{Lind} {et~al.}(2011){Lind}, {Charbonnel}, {Decressin}, {Primas},
  {Grundahl}, \& {Asplund}}]{Lind11}
{Lind}, K., {Charbonnel}, C., {Decressin}, T., {Primas}, F., {Grundahl}, F., \&
  {Asplund}, M. 2011, \aap, 527, A148

\bibitem[{{Marigo} {et~al.}(2008){Marigo}, {Girardi}, {Bressan}, {Groenewegen},
  {Silva}, \& {Granato}}]{Marigo08}
{Marigo}, P., {Girardi}, L., {Bressan}, A., {Groenewegen}, M.~A.~T., {Silva},
  L., \& {Granato}, G.~L. 2008, \aap, 482, 883

\bibitem[{{Narayanan} \& {Dav{\'e}}(2012)}]{Narayanan12}
{Narayanan}, D. \& {Dav{\'e}}, R. 2012, \mnras, 423, 3601

\bibitem[{{Oconnell}(1976)}]{Oconnell76}
{Oconnell}, R.~W. 1976, \apj, 206, 370

\bibitem[{{Oser} {et~al.}(2012){Oser}, {Naab}, {Ostriker}, \&
  {Johansson}}]{Oser12}
{Oser}, L., {Naab}, T., {Ostriker}, J.~P., \& {Johansson}, P.~H. 2012, \apj,
  744, 63

\bibitem[{{Padoan} \& {Nordlund}(2002)}]{Padoan02}
{Padoan}, P. \& {Nordlund}, {\AA}. 2002, \apj, 576, 870

\bibitem[{{Padoan} {et~al.}(1997){Padoan}, {Nordlund}, \& {Jones}}]{Padoan97}
{Padoan}, P., {Nordlund}, A., \& {Jones}, B.~J.~T. 1997, \mnras, 288, 145

\bibitem[{{Rayner} {et~al.}(2009){Rayner}, {Cushing}, \& {Vacca}}]{Rayner09}
{Rayner}, J.~T., {Cushing}, M.~C., \& {Vacca}, W.~D. 2009, \apjs, 185, 289

\bibitem[{{Saglia} {et~al.}(2010)}]{Saglia10}
{Saglia}, R.~P. {et~al.} 2010, \aap, 509, A61

\bibitem[{{S{\'a}nchez-Bl{\'a}zquez} {et~al.}(2006){S{\'a}nchez-Bl{\'a}zquez},
  {Peletier}, {Jim{\'e}nez-Vicente}, {Cardiel}, {Cenarro},
  {Falc{\'o}n-Barroso}, {Gorgas}, {Selam}, \& {Vazdekis}}]{Sanchez-Blazquez06}
{S{\'a}nchez-Bl{\'a}zquez}, P., {Peletier}, R.~F., {Jim{\'e}nez-Vicente}, J.,
  {Cardiel}, N., {Cenarro}, A.~J., {Falc{\'o}n-Barroso}, J., {Gorgas}, J.,
  {Selam}, S., \& {Vazdekis}, A. 2006, \mnras, 371, 703

\bibitem[{{Scalo}(1986)}]{Scalo86}
{Scalo}, J.~M. 1986, Fundamentals of Cosmic Physics, 11, 1

\bibitem[{{Schiavon}(2007)}]{Schiavon07}
{Schiavon}, R.~P. 2007, \apjs, 171, 146

\bibitem[{{Schiavon} {et~al.}(2005){Schiavon}, {Rose}, {Courteau}, \&
  {MacArthur}}]{Schiavon05}
{Schiavon}, R.~P., {Rose}, J.~A., {Courteau}, S., \& {MacArthur}, L.~A. 2005,
  \apjs, 160, 163

\bibitem[{{Scott} {et~al.}(2009)}]{Scott09}
{Scott}, N. {et~al.} 2009, \mnras, 398, 1835

\bibitem[{{Scwarz}(1978)}]{Schwarz78}
{Scwarz}, G. 1978, Ann. Statist., 5, 461

\bibitem[{{Silk}(1995)}]{Silk95}
{Silk}, J. 1995, \apjl, 438, L41

\bibitem[{{Smith {et~al.}}(2012)}]{Smith12b}
{Smith}, R.~J., {Lucey}, J.~R., {Carter}, D. 2012, \mnras, 421, 2982

\bibitem[{{Sonnenfeld} {et~al.}(2012){Sonnenfeld}, {Treu}, {Gavazzi},
  {Marshall}, {Auger}, {Suyu}, {Koopmans}, \& {Bolton}}]{Sonnenfeld12}
{Sonnenfeld}, A., {Treu}, T., {Gavazzi}, R., {Marshall}, P.~J., {Auger}, M.~W.,
  {Suyu}, S.~H., {Koopmans}, L.~V.~E., \& {Bolton}, A.~S. 2012, \apj, 752, 163

\bibitem[{{Spiniello} {et~al.}(2012){Spiniello}, {Trager}, {Koopmans}, \&
  {Chen}}]{Spiniello12}
{Spiniello}, C., {Trager}, S.~C., {Koopmans}, L.~V.~E., \& {Chen}, Y.~P. 2012,
  \apjl, 753, L32

\bibitem[{{Spinrad}(1962)}]{Spinrad62}
{Spinrad}, H. 1962, \apj, 135, 715

\bibitem[{{Spinrad} \& {Taylor}(1971)}]{Spinrad71}
{Spinrad}, H. \& {Taylor}, B.~J. 1971, \apjs, 22, 445

\bibitem[{{Strader} {et~al.}(2011){Strader}, {Caldwell}, \& {Seth}}]{Strader11}
{Strader}, J., {Caldwell}, N., \& {Seth}, A.~C. 2011, \aj, 142, 8

\bibitem[{{Thomas} {et~al.}(2003){Thomas}, {Maraston}, \& {Bender}}]{Thomas03}
{Thomas}, D., {Maraston}, C., \& {Bender}, R. 2003, \mnras, 339, 897

\bibitem[{{Thomas} {et~al.}(2005){Thomas}, {Maraston}, {Bender}, \& {Mendes de
  Oliveira}}]{Thomas05}
{Thomas}, D., {Maraston}, C., {Bender}, R., \& {Mendes de Oliveira}, C. 2005,
  \apj, 621, 673

\bibitem[{{Thomas} {et~al.}(2011{\natexlab{a}}){Thomas}, {Maraston}, \&
  {Johansson}}]{Thomas11}
{Thomas}, D., {Maraston}, C., \& {Johansson}, J. 2011{\natexlab{a}}, \mnras,
  412, 2183

\bibitem[{{Thomas} {et~al.}(2011{\natexlab{b}}){Thomas}, {Saglia}, {Bender},
  {Thomas}, {Gebhardt}, {Magorrian}, {Corsini}, {Wegner}, \&
  {Seitz}}]{ThomasJ11}
{Thomas}, J., {Saglia}, R.~P., {Bender}, R., {Thomas}, D., {Gebhardt}, K.,
  {Magorrian}, J., {Corsini}, E.~M., {Wegner}, G., \& {Seitz}, S.
  2011{\natexlab{b}}, \mnras, 415, 545

\bibitem[{{Trager} {et~al.}(2000){Trager}, {Faber}, {Worthey}, \&
  {Gonz{\'a}lez}}]{Trager00}
{Trager}, S.~C., {Faber}, S.~M., {Worthey}, G., \& {Gonz{\'a}lez}, J.~J. 2000,
  \aj, 120, 165

\bibitem[{{Treu} {et~al.}(2010){Treu}, {Auger}, {Koopmans}, {Gavazzi},
  {Marshall}, \& {Bolton}}]{Treu10}
{Treu}, T., {Auger}, M.~W., {Koopmans}, L.~V.~E., {Gavazzi}, R., {Marshall},
  P.~J., \& {Bolton}, A.~S. 2010, \apj, 709, 1195

\bibitem[{{van Dokkum} \& {Conroy}(2010)}]{vanDokkum10}
{van Dokkum}, P.~G. \& {Conroy}, C. 2010, \nat, 468, 940

\bibitem[{{van Dokkum} \& {Conroy}(2011)}]{vanDokkum11}
---. 2011, \apjl, 735, L13

\bibitem[{{Whitford}(1977)}]{Whitford77}
{Whitford}, A.~E. 1977, \apj, 211, 527

\bibitem[{{Worthey} {et~al.}(1992){Worthey}, {Faber}, \&
  {Gonzalez}}]{Worthey92}
{Worthey}, G., {Faber}, S.~M., \& {Gonzalez}, J.~J. 1992, \apj, 398, 69

\end{thebibliography}



\appendix
\setcounter{figure}{0}
\renewcommand*\thefigure{A\arabic{figure}}

In an online-only appendix we provide the fits to all galaxies
included in Paper I (excluding NGC 4621 and NGC 524, which are shown
in Figures 1 and 2 in the main text).

\end{document}